\newcommand\vpy[3]{{\bf #1}, #2 (#3)}

\newcommand\nat[3]{{\rm Nature~\vpy{#1}{#2}{#3}}} 
 
\newcommand\prl[3]{{\rm Phys.~Rev.~Lett.~\vpy{#1}{#2}{#3}}} 
\newcommand\pr[3]{{\rm Phys.~Rev.~\vpy{#1}{#2}{#3}}}

\newcommand\apj[3]{{\rm Ap.~J.~\vpy{#1}{#2}{#3}}}
\newcommand\mnras[3]{{\rm Mon.~Not.~R.~Astron.~Soc.~\vpy{#1}{#2}{#3}}}
\newcommand\araa[3]{{\rm Annu.~Rev.~Astron.~Astrophys.~\vpy{#1}{#2}{#3}}}

\newcommand\astroph[2]{{\it `#1'} {\rm (see astro-ph/#2)}}

\documentstyle[prl,aps,epsf,floats,eqsecnum]{revtex}

\begin{document}
\title{Cosmological matter perturbations}

\author{
  Jiun-Huei Proty Wu
  }
\address{
  Astronomy Department,
  University of California, Berkeley,\\
  601 Campbell Hall, Berkeley, CA 94720-3411, USA
}

\maketitle

\begin{abstract}

We investigate matter density perturbations
in models of structure formation with or without causal/acausal source.
Under the fluid approximation in the linear theory,
we first derive full perturbation equations
in flat space with a cosmological constant $\Lambda$.
We then use Green-function technique
to obtain analytic solutions for matter perturbations
in a flat $\Lambda=0$ model.
Some incorrect solutions in the literature are corrected here.
A simple yet accurate extrapolation scheme
is then proposed to obtain solutions in curved or $\Lambda\neq 0$ cosmologies.
Some general features of these solutions are revealed.
In particular,
  we analytically prove that
  the resulting matter density perturbations are independent of the way 
  the causal source was compensated into the background contents
  of the universe when it was first formed.
We also use our Green-function solutions to investigate
  the compensation mechanism for perturbations with causal seeds,
  and yield a mathematically and physically explicit form
  in interpreting it.
  We found that
  the compensation scale depends 
not only on the dynamics of the universe,
but also on the properties of the seeds near the horizon scale.
It can be accurately located by employing our Green functions.

\end{abstract}
\vskip .2in


\section{Introduction}

The standard cosmology was lack of a mechanism to produce cosmological 
perturbations.
In order to compensate for this flaw in the standard model,
there are currently two main paradigms for 
structure formation---inflation \cite{Guth}
and topological defects \cite{VilShe}.
While
the beauty and simplicity of the former appears to have
enticed more adherents and studies,
the latter has proved computationally
much more challenging to make robust predictions with which
to confront observations
\cite{AveShe2,AveShe4,AveShe3,PST,against,ABR2,ConHin,AveCal1,ACDKSS,ruth}.
These two paradigms are fundamentally different
in the way they generate cosmological perturbations.
The standard adiabatic inflation produces primordial perturbations 
on all scales of cosmological interest
via quantum fluctuations
and the causal constraint during inflation,
and these perturbations grow over time
in an uncorrelated manner.
As a consequence,
the perturbations today can be thought of as
simply transfered from the initial irregularities that inflation set up,
and this transfer function can be easily obtained in the linear theory
and thus well understood in the literature.
On the other hand,
topological defects are the byproducts
of the spontaneous symmetry-breaking phase transition
in the early universe,
and hence carry energy that was carved out of 
the originally homogeneous background energy of the universe.
Therefore due to causality,
defects induce perturbations only on sub-horizon scales,
via gravitational interactions while evolving.
This mechanism that prevents the growth of super-horizon perturbations
is called the `compensation mechanism'.
In addition,
due to the certain topology of the defect network,
the resulting perturbations are correlated and thus non-Gaussian,
in contrast to the standard adiabatic inflationary perturbations.
It then follows that
to compute the perturbations in models with defects,
we need to know the evolution history of defects
for the entire dynamic range 
during which the cosmological perturbations of our interest were seeded.
This is what makes 
the computation of defect-induced perturbations so difficult.

In the literature
the power spectra of this kind of models have been investigated
using the full Einstein-Boltzmann equations.
However,
the study of the phase information of these perturbations 
still remains difficult because of the limited computation power.
Although
there have been some detailed treatments for theories with causal seeds
\cite{HuSelWhi,HuWhi},
we shall in this paper present a simpler formalism, 
which is an approximation to the full Einstein-Boltzmann equations,
to provide not only a physically transparent way for understanding
the evolution of density perturbations in models with source,
but also a computationally economical scheme
to investigate the phase information
of the resulting cosmological perturbations. 
This formalism is parallel to
those presented in Ref.~\cite{VeeSte} and Ref.~\cite{turok},
but 
we give some modifications
to incorporate
the inclusion of the cosmological constant and
a more detailed treatment for the effect of baryon-photon coupling/decoupling.
We also note that
part of the solutions in Ref.~\cite{VeeSte} are incorrect 
due to the incorrect initial conditions and
the incorrect assumptions about the form of the subsequent perturbations
induced by the source (see text later).
We shall correct these mistakes
and further provide a complete and explicit set of analytic solutions
for the matter density perturbations.
With an accurate extrapolation scheme,
these solutions become also valid for models
with any reasonably chosen background cosmology.
The formalism and its solutions to be developed here
will be completely general and thus suitable for any models
with or without causal/acausal source.

The structure of this paper is as follows.
In section~\ref{synchronous_gauge_perturbation_theory},
under the fluid approximation,
we first derive in the synchronous gauge
the full perturbation equations with source terms,
in flat cosmologies with a cosmological constant $\Lambda$.
This is done by considering the stress-energy conservation of 
the fluids (\ref{conservation-rad-mat}) 
and the source (\ref{conservation-seeds}),
and the linearly perturbed Einstein equations (\ref{perturbed-einstein}).
The fluid components considered here are
cold dark matter (CDM), baryons (B), and photons ($\gamma$),
and we employ  the baryon-photon tight-coupling approximation
to derive the perturbation equations before the last-scattering epoch.
In this context,
we also investigate the role of the so-called stress-energy pseudotensor
(\ref{conservation-pseudo}).
The initial conditions of these perturbation equations 
are discussed (\ref{initial-conditions}),
and we use the approximation of instantaneous decoupling 
to deal with the decoupling of
photons and baryons at the epoch of last scattering
(\ref{instantaneous-decoupling}).
We then numerically justify the accuracy of this formalism
in the context of standard CDM models,
by
comparing its results with those of 
the full Einstein-Boltzmann solver \cite{cmbfast} (\ref{verification-stdcdm}).
Within reasonable ranges of cosmological parameters,
our approach provides satisfactory precision
at greatly reduced numerical cost.

In section~\ref{solutions-of-matter-perturbations},
we derive the matter perturbation solutions of the equations 
presented in section~\ref{synchronous_gauge_perturbation_theory}.
The perturbations of radiation and matter are first divided into two parts:
the initial and the subsequent perturbations.
With some change of variables,
these equations are then ready to be solved by the Green-function technique 
(\ref{evolution-equations-and-decomposition-of-perturbation}).
With this technique,
we find the exact solutions on scales much larger or much smaller 
than the horizon size,
namely the super-horizon or the sub-horizon solutions respectively
(\ref{super-horizon-and-sub-horizon-modes}).
Some degeneracy among the Green functions 
for the matter perturbation solutions is then found
and used to reduce their effective number (\ref{degeneracy-of-the-green-functions}).
With this great simplification,
solutions on intermediate scales are then easily obtained
by an accurate interpolation scheme
based on the well-known standard CDM transfer function
(\ref{solutions-on-intermediate-scales}).
We also discuss the effect of baryons (\ref{the-effect-of-baryons}).
A simple and accurate extrapolation scheme 
is then introduced to obtain solutions
in the $K\neq 0$ or $\Lambda\neq 0$ cosmologies
(\ref{k-neq-0-or-l-neq-0-models}),
where $K$ is the curvature of the universe
 (see Appendix~\ref{background_cosmologies}).
All our Green-function solutions are numerically verified to high accuracy.

In section~\ref{important-properties},
we use our Green-function solutions to
investigate some important properties
of cosmological matter density perturbations.
We first demonstrate
the relation between our solutions and the standard CDM transfer function
(\ref{the-standard-cdm-model}).
We also prove that in models with causal source,
the resulting matter perturbations today are independent of
the way the source energy is initially compensated into the background
contents of the universe
(\ref{independence-of-the-initial-conditions}).
Finally we use our Green-function solutions to study
the compensation mechanism and the scale on which it operates
(\ref{compensation-and-total-matter-perturbations}).
We find that this compensation scale is determined
not only by the dynamics of the universe,
but also by the properties of the source near the horizon scale.
Once the detailed features of the source near the horizon scale are known,
this compensation scale can be accurately located
using our Green functions.
A summary and conclusion is  given in section~\ref{matpert-conclusion}.
In appendix~\ref{background_cosmologies},
we define the convention of some notations used in this paper,
and
present for reference
the solutions for the dynamics of various background cosmologies,
including the consideration of non-zero curvature and a cosmological constant.


\section{Synchronous gauge perturbation theory}
\label{synchronous_gauge_perturbation_theory}

In this section,
we derive the linear evolution equations for cosmological perturbations.
To calculate the density and metric perturbations,
we model the contents of the universe as perfect fluids:
radiation (photons and neutrinos) and 
pressureless matter (CDM and baryons).
We shall use the photon-baryon tight-coupling approximation until the epoch
of last scattering, 
at which we assume instantaneous decoupling,
also taking into account the effect of Silk damping 
due to the photon diffusion.
After the decoupling,
the baryonic perturbations originating from the perturbations
of the photon-baryon coupled fluid
are then merged linearly into the CDM content.
In scenarios with causal seeds,
the radiation and matter fields are assumed to be initially uniform,
and then perturbed by the causal seeds after they are formed.
The radiation, matter, and causal seeds are assumed to interact only
through gravity,
meaning that their stress-energy tensors are separately covariantly conserved.

We shall work in the synchronous gauge,
in which the perturbations $h_{\mu\nu}$ to the spacetime metric $g_{\mu\nu}$
obey the constraint $h_{0\mu}=0$.
Throughout this paper,
we use a signature $(-+++)$ for the spacetime metric,
and units in which $\hbar=c=k_{\rm B}=1$.
Thus the perturbed flat Friedmann-Robertson-Walker (FRW) metric is given by
\begin{equation}
  g_{00} = -a^2(\eta), \quad 
  g_{ij} =  a^2(\eta)[\delta_{ij}+h_{ij}(\eta, {\bf x})].
\end{equation}
We shall work in the linear theory, requiring $|h_{ij}| \ll 1$.
Greek alphabet will denote the spacetime indices (e.g.\ $\mu=0,1,2,3$),
and mid-alphabet Latin letters the spatial indices (e.g.\ $i=1,2,3$).
Although the synchronous gauge is sometimes criticised in the literature
due to its residual gauge freedom,
it is still well suited to models in which 
the universe evolves from being perfectly homogeneous and isotropic.
In such models,
all perturbation variables can be initially set to zero
(before the causal seeds are generated),
and this is normally referred to as the
`initially unperturbed synchronous gauge' (IUSG) \cite{VeeSte}.
It possesses no residual gauge freedom.
Thus the Einstein equations are completely causal in IUSG,
with the values of all perturbation variables at a given spacetime point
being completely determined by initial conditions
within the past light cone of the point.
One example of such models is the cosmic defect models,
which have been of most interest in the study of models with causal seeds.

In section~\ref{conservation-rad-mat},
we derive in the IUSG the conservation equations of 
radiation and matter fields.
In section~\ref{conservation-seeds},
we consider the conservation of source stress energy.
In section~\ref{perturbed-einstein},
we derive the linearly perturbed Einstein equations.
Then, in section~\ref{conservation-pseudo},
we employ the concept of stress-energy pseudotensor to
investigate the internal energy transfer among various fields.
In section~\ref{instantaneous-decoupling},
we describe the approximation of instantaneous decoupling of
photons and baryons at the epoch of last scattering.
In section~\ref{verification-stdcdm},
we numerically verify the accuracy of our formalism for the standard CDM model,
in comparison with the results from CMBFAST \cite{cmbfast}, 
a fast Einstein-Boltzmann solver.

\subsection{Stress-energy conservation of radiation and matter fields}
\label{conservation-rad-mat}

The contents of the universe are considered as perfect fluids,
whose energy-momentum tensors have the form
\begin{equation}
  \label{T-munu}
  T^{\mu}_{{\scriptscriptstyle N}\nu}=(\rho_{\scriptscriptstyle N}+p_{\scriptscriptstyle N}) u_{\scriptscriptstyle N}^\mu u_{{\scriptscriptstyle N}\nu} + p_{\scriptscriptstyle N} \delta^\mu_\nu,
  \quad
  \textrm{with }
  u_{\scriptscriptstyle N}^\mu u_{{\scriptscriptstyle N}\mu}=-1.
\end{equation}
Here $\rho_{\scriptscriptstyle N}$, $p_{\scriptscriptstyle N}$, and $u_{\scriptscriptstyle N}^\mu$ are the density, pressure,
and four-velocity of the $N$th fluid respectively.
In the homogeneous background,
we have $u_{\scriptscriptstyle N}^\mu=(a^{-1}, \vec{0})$,
which implies that $\delta u_{\scriptscriptstyle N}^0=0$ to first order.
We thus define the velocity perturbation as 
$v_{\scriptscriptstyle N}^i=a\delta u_{\scriptscriptstyle N}^i$,
i.e., $\delta u^\mu_{\scriptscriptstyle N}=(0,{\bf v}_{\scriptscriptstyle N}/a)$.
The equation of state and the sound speed are defined respectively as
\begin{equation}
  \label{eqn-of-state}
  \kappa_{\scriptscriptstyle N} = \frac{p_{\scriptscriptstyle N}}
  {\rho_{\scriptscriptstyle N}},  
  \quad
  c_{\scriptscriptstyle N}^2=\frac{\delta p_{\scriptscriptstyle N}}
  { \delta \rho_{\scriptscriptstyle N}}.
\end{equation}
Consequently, the covariant conservation of stress energy for each fluid
$T^{\mu\nu}_{{\scriptscriptstyle N};\nu}=0$ gives \cite{VeeSte}
\begin{eqnarray}
  \dot{\delta}_{\scriptscriptstyle N}+ (1+\kappa_{\scriptscriptstyle N})
  (\nabla\cdot {\bf v}_{\scriptscriptstyle N}+\frac{1}{2}\dot{h})
  +3\frac{\dot{a}}{a}(c_{\scriptscriptstyle N}^2-\kappa_{\scriptscriptstyle N}) \delta_{\scriptscriptstyle N}=0,
  \label{conserve-1}\\
  \dot{\bf v}_{\scriptscriptstyle N}+ \frac{\dot{a}}{a}(1-3c_{\scriptscriptstyle N}^2){\bf v}_{\scriptscriptstyle N}
  + \frac{c_{\scriptscriptstyle N}^2}{1+\kappa_{\scriptscriptstyle N}} \nabla\delta_{\scriptscriptstyle N}=0,
  \label{conserve-2}\\
  \dot{\bf v}_{\scriptscriptstyle N}^{\perp}+\frac{\dot{a}}{a}(1-3c_{\scriptscriptstyle N}^2){\bf v}^{\perp}_{\scriptscriptstyle N}=0,
  \label{conserve-3}
\end{eqnarray}
where 
$\delta_{\scriptscriptstyle N}=\delta \rho_{\scriptscriptstyle N}/\rho_{\scriptscriptstyle N}$,
$h\equiv h_{ii}$ is the spatial trace of $h_{\mu\nu}$,
and
we have decomposed the velocities as 
${\bf v}_{\scriptscriptstyle N}={\bf v}_{\scriptscriptstyle N}^\parallel+{\bf v}_{\scriptscriptstyle N}^\perp$, 
with $\nabla\times {\bf v}_{\scriptscriptstyle N}^\parallel=0$ and $\nabla\cdot {\bf v}_{\scriptscriptstyle N}^\perp=0$.

In the regime of photon-baryon tight coupling,
we have only two main fluids:
the CDM component and the tightly-coupled photon-baryon fluid.
They will be denoted as $N={\rm c}, \gamma {\rm B}$ respectively,
and discussed separately as follows.
Note that we have ignored the neutrinos in the radiation.

\subsubsection{CDM fluid}
We first consider the CDM fluid, i.e.\ $N={\rm c}$.
With $\kappa_{\rm c}=c_{\rm c}^2=0$ for pressureless matter,
the equations of stress-energy conservation 
(\ref{conserve-1})--(\ref{conserve-3}) become
\begin{equation}
  \dot{\delta}_{\rm c}+ \nabla\cdot {\bf v}_{\rm c}= -\frac{1}{2}\dot{h},  
  \quad
  \dot{\bf v}_{\rm c}+\frac{\dot{a}}{a} {\bf v}_{\rm c}=0.
  \label{conserve-c0}
\end{equation}
As we can see,
any perturbations in the CDM velocity will decay as $a^{-1}$.
Thus we can simply choose ${\bf v}_{\rm c}=\vec{0}$ in the IUSG.
Once ${\bf v}_{\rm c}=\vec{0}$, it will remain so
as there is no linear gravitational source.
As a consequence, 
the CDM obeys a single nontrivial conservation law resulting from
equation (\ref{conserve-c0})
\begin{equation}
  \label{conserve-c}
  \dot{h}+2\dot{\delta_{\rm c}}=0
  \Longrightarrow
  h = -2 \delta_{\rm c},
\end{equation}
where the second equation results from 
the initial condition $h=\delta_{\rm c}=0$,
as required by the IUSG.

\subsubsection{Photon-baryon tightly coupled fluid and its photon component}

For the  tightly-coupled photon-baryon ($\gamma \rm B$) fluid,
we have
\begin{equation}
  \label{v-p-rho-gammaB}
  {\bf v}_{\gamma \rm B}={\bf v}_{\gamma}={\bf v}_{\rm B},
  \quad
  p_{\gamma \rm B}=p_{\gamma},
  \quad
  \rho_{\gamma \rm B}=\rho_{\gamma}+\rho_{\rm B}\,.
\end{equation}
Thus we can define 
\begin{equation}
  \label{R-gammaB}
  R
  =\frac{\delta\rho_{\rm B}}{\delta\rho_\gamma}
  = \frac{3\rho_{\rm B}}{4\rho_{\gamma}},
\end{equation}
where the second result comes from the fact that 
$\rho_{\gamma}\propto a^{-4}$ and $\rho_{\rm B}\propto a^{-3}$.
Definitions (\ref{eqn-of-state}) then give
\begin{equation}
  \label{c_gammaB}
  \kappa_{\gamma {\rm B}}=\frac{1}{3+4R},
  \quad
  c_{\gamma \rm B}^2= \frac{1}{3(1+R)}.
\end{equation}
With these results,
the equations of stress-energy conservation for the $\gamma \rm B$ fluid
can be obtained from equations (\ref{conserve-1})--(\ref{conserve-3}):
\begin{eqnarray}
  \dot{\delta}_{\gamma {\rm B}}
  +\frac{4+4R}{3+4R}(\nabla\cdot {\bf v}_{\gamma {\rm B}}-\dot{\delta}_{\rm c})+
  \frac{\dot{a}}{a}\frac{R}{(1+R)(3+4R)}\delta_{\gamma {\rm B}}=0, \\
  \dot{\bf v}_{\gamma {\rm B}}
  + \frac{\dot{a}}{a}\frac{R}{1+R} {\bf v}_{\gamma {\rm B}}+ 
  \frac{3+4R}{12(1+R)^2} \nabla  \delta_{\gamma {\rm B}}=0,\\
  \dot{\bf v}_{\gamma \rm B}^{\perp}
  +\frac{\dot{a}}{a}\frac{R}{1+R} {\bf v}^{\perp}_{\gamma \rm B}=0.
\end{eqnarray}
In cosmological applications, such as CMB anisotropies,
we are more interested in the photon perturbations
rather than the perturbations in the $\gamma B$ fluid.
Therefore by using equations (\ref{v-p-rho-gammaB}) and (\ref{R-gammaB}),
we can  extract the photon component from the above equations
to yield \cite{neils-idea-gammaB}
\begin{eqnarray}
  \dot{\delta}_{{\rm r}}+\frac{4}{3} \nabla\cdot {\bf v}_{{\rm r}}
  -\frac{4}{3}{\dot{\delta}}_{\rm c}=0,
  \label{conserve-r1}\\
  \dot{\bf v}_{{\rm r}}+ \frac{\dot{a}}{a}\frac{R}{1+R} {\bf v}_{{\rm r}}
  + \frac{1}{4+4R} \nabla \delta_{{\rm r}}=0,
  \label{conserve-r2}
\end{eqnarray}
where we have ignored neutrinos in the radiation
so as to replace the subscript $\gamma$ with $\rm r$.
The velocity can then be eliminated to yield
a single second-order equation:
\begin{equation}
  \label{conserve-r}
  \ddot{\delta}_{{\rm r}}-\frac{4}{3}\ddot{\delta}_{\rm c}
  +\frac{\dot{R}}{1+R}(\dot{\delta}_{{\rm r}}-\frac{4}{3}\dot{\delta}_{\rm c})
  -\frac{1}{3+3R}\nabla^2 \delta_{{\rm r}}=0.
\end{equation}
We note that
although the photon velocities are missing in this equation,
they can be recovered at any given moment using equation~(\ref{conserve-r1}).

An alternative presentation of equations (\ref{conserve-r1}) 
and (\ref{conserve-r}) is via the entropy perturbation $s$.
It is defined as the fluctuation 
in the number of photons per dark matter particle
\begin{equation}
  \label{entropy}
  s=\frac{3}{4}\delta_{{\rm r}}-\delta_{\rm c}.
\end{equation}
Thus equations (\ref{conserve-r1}) and (\ref{conserve-r}) can be
rewritten as
\begin{eqnarray}
  \dot{s} 
  & = & - \nabla \cdot {\bf v}_{{\rm r}},
  \label{entropy1}\\
  \ddot{s}
  & = &
  - \frac{\dot{R}}{1+R}\dot{s}
  + \frac{1}{3+3R}\nabla^2(s+\delta_{\rm c})\;.
  \label{entropy2}
\end{eqnarray}
As we shall see, 
$\delta_{\rm r}$ can only have a white noise power spectrum on super-horizon scales.
From equation (\ref{conserve-r2}),
this implies a $k^2$ power spectrum in ${\bf v}_{\rm r}$ on these scales.
Adding the fact that the entropy fluctuation $s$ starts from zero
on super-horizon scales due to 
the fixed number of dark matter particles per photon,
it then follows from equation (\ref{entropy1}) that
both $s$ and $\dot{s}$ have a $k^4$ fall off outside the horizon.
Therefore in numerical simulations,
as long as the initial horizon size is smaller than the scales of our interest,
we can simply set $s=\dot{s}=0$ as part of the initial condition.

\subsection{Stress-energy conservation of the source}
\label{conservation-seeds}

The causal source we shall consider is weak, 
so it will appear only as first-order terms 
in the perturbed Einstein equations.
Thus in the linear theory we are considering here,
they can be treated as being stiff,
meaning that their evolution depends only on their own self-interactions
and the background dynamics of the universe,
but not on their self-gravity 
or on the weak gravitational field of the 
inhomogeneities they produce.
This assumption will enable us to separate the calculation of their dynamics
from that of the inhomogeneities they induce,
allowing us to evolve them
as if they are in a completely homogeneous background.
Since the source is stiff,  
its energy-momentum tensor $\Theta_{\mu\nu}$ need only be 
locally covariantly conserved with respect to the background:
\begin{eqnarray}
  \Theta_{00,0}+ \frac{\dot{a}}{a}\Theta_+ & = & \Theta_{0i,i} \;,
  \label{conserve-seed1}\\
  \Theta_{0i,0}+ 2\frac{\dot{a}}{a}\Theta_{0i} & = & \Theta_{ij,j} \;,
  \label{conserve-seed2}
\end{eqnarray}
where $\Theta_{+}=\Theta_{00}+\Theta_{ii}$.

Another important aspect of 
cosmic structure formation with causal seeds like cosmic defects 
is the fact that
the sources, formed at very early times,
will ultimately create under-densities 
in the initially homogeneous background, out of which they are carved.
This is a direct result of energy conservation in the universe,
and is normally termed `compensation'.
We shall discuss this issue in more detail later.

\subsection{Linearly Perturbed Einstein equations}
\label{perturbed-einstein}

At first we have ten Einstein equations
\begin{equation}
  \label{Einstein00}
  R_{\mu\nu} =
  8\pi G(T_{\mu\nu}-\frac{1}{2}g_{\mu\nu}T) + \Lambda g_{\mu\nu},
\end{equation}
or equivalently,
\begin{equation}
  \label{Einstein0}
  G_{\mu\nu}\equiv
  R_{\mu\nu}-\frac{1}{2}g_{\mu\nu}R_{\rm S} =
  8\pi GT_{\mu\nu} - \Lambda g_{\mu\nu},
\end{equation}
where 
$R_{\mu\nu}$ is the Ricci tensor,
$G$ is the gravitational constant,
$T=g^{\mu\nu}T_{\mu\nu}$,
$\Lambda$ is the cosmological constant,
$G_{\mu\nu}$ is the Einstein tensor,
and 
$R_{\rm S}\equiv g^{\mu\nu}R_{\mu\nu}$ is the scalar curvature.
Linearly perturbing the above equations, we obtain
\begin{equation}
  \label{dEinstein00}
  \delta R_{\mu\nu}=8\pi G(
    \delta T_{\mu\nu}
    -\frac{1}{2}h_{\mu\nu}\eta^{rs}T_{rs}
    -\frac{1}{2}\eta_{\mu\nu}\eta^{rs}\delta T_{rs}
    +\frac{1}{2}\eta_{\mu\nu}h_{pq}\eta^{rp}\eta^{sq}T_{rs})
  +a^2\Lambda h_{\mu\nu},
\end{equation}
or equivalently,
\begin{equation}
  \label{dEinstein0}
  \delta G_{\mu\nu}=8\pi G \delta T_{\mu\nu} - a^2\Lambda h_{\mu\nu},
\end{equation}
where
\begin{equation}
  \label{deltaT}
  \delta T_{\mu\nu}=\Theta_{\mu\nu}
      + a^2\sum_N (h_{\mu r}T^r_{N\nu} + \eta_{\mu s}\delta T^s_{N\nu}).
\end{equation}
A closed set of the ten linearly perturbed Einstein equations are then
\begin{eqnarray}
  2\delta R_{00} 
  =  -\ddot{h}-{\frac{\dot{a}}{a}}\dot{h}
  & = &
  +3\left({\frac{\dot{a}}{a}}\right)^2
  \sum_N (1+3c_{\scriptscriptstyle N}^2)\Omega_N\delta_N
  +8\pi G \Theta_+ \;,
  \label{dEinstein1}\\
  2\left\{\delta \tilde{R}_{ij}-\left[{\frac{\ddot{a}}{a}}+
  \left({\frac{\dot{a}}{a}}\right)^2\right] \tilde{h}_{ij}\right\}
  & = &
  \ddot{\tilde{h}}_{ij}+2{\frac{\dot{a}}{a}}\dot{\tilde{h}}_{ij}
  -\nabla^2 {\tilde{h}}_{ij}-{\frac{1}{3}}h_{,ij}
  +{\frac{1}{9}}\delta_{ij}\nabla^2 h
  \nonumber\\
  & & 
  +\tilde{h}_{ik,kj}
  +\tilde{h}_{jk,ki}-{\frac{2}{3}}\delta_{ij}\tilde{h}_{kl,kl}
  =16\pi G\tilde{\Theta}_{ij} + a^2 \Lambda \tilde{h}_{ij}\;,
  \label{dEinstein2}\\
  2\delta G_{00} 
  = \tilde{h}_{ij,ij}-{\frac{2}{3}}\nabla^2h+2{\frac{\dot{a}}{a}}\dot{h}
  & = &
  6\left({\frac{\dot{a}}{a}}\right)^2
  \sum_N \Omega_N\delta_N
  +16\pi G\Theta_{00}\;,
  \label{dEinstein3}\\
  2\delta G_{0i}
  =  \dot{\tilde{h}}_{ij,j}-{\frac{2}{3}}\dot{h}_{,i}
  & = &
  -6\left({\frac{\dot{a}}{a}}\right)^2
  {\sum_N}' (1+\kappa_{\scriptscriptstyle N})\Omega_N v^i_{\scriptscriptstyle N}
  +16\pi G\Theta_{0i}\;,
  \label{dEinstein4}
\end{eqnarray}
where the traceless parts are defined by 
$\tilde{R}_{ij}=R_{ij}-\delta_{ij}{R^k}_k/3$, and similarly for 
$\tilde{h}_{ij}$ and $\tilde{\Theta}_{ij}$.
The prime over the sum in equation (\ref{dEinstein4}) indicates the sum
over all fluids except CDM.
We note from the above results that in the IUSG
the cosmological constant $\Lambda$ does not appear as extra terms in
the perturbation equations except in (\ref{dEinstein2}), the `$ij$' component.

Within the photon-baryon tight-coupling regime,
the above perturbation equations simplify as:
\begin{eqnarray}
  -\ddot{h}-{\frac{\dot{a}}{a}}\dot{h}
  =
  +3\left({\frac{\dot{a}}{a}}\right)^2
  \left[(2+R)\Omega_{{\rm r}}\delta_{\rm r}+\Omega_{\rm c}\delta_{\rm c}\right]+
  8\pi G \Theta_+ \;,
  \label{Einstein1}\\
  \ddot{\tilde{h}}_{ij}+2{\frac{\dot{a}}{a}}\dot{\tilde{h}}_{ij}
  -\nabla^2 {\tilde{h}}_{ij}-{\frac{1}{3}}h_{,ij}
  +{\frac{1}{9}}\delta_{ij}\nabla^2 h
  +\tilde{h}_{ik,kj}
  +\tilde{h}_{jk,ki}-{\frac{2}{3}}\delta_{ij}\tilde{h}_{kl,kl}
  =16\pi G\tilde{\Theta}_{ij} + a^2 \Lambda \tilde{h}_{ij}\;,
  \label{Einstein2}\\
  \tilde{h}_{ij,ij}-{\frac{2}{3}}\nabla^2h+2{\frac{\dot{a}}{a}}\dot{h}
  =
  6\left({\frac{\dot{a}}{a}}\right)^2
  \left[\Omega_{\rm c}\delta_{\rm c}+(1+R)\Omega_{\rm r}\delta_{\rm r}\right]
  +16\pi G\Theta_{00}\;,
  \label{Einstein3}\\
  \dot{\tilde{h}}_{ij,j}-{\frac{2}{3}}\dot{h}_{,i}
  =
  -8\left({\frac{\dot{a}}{a}}\right)^2
  (1+R)\Omega_{\rm r} v_{\rm r}^i+16\pi G\Theta_{0i}\;.
  \label{Einstein4}
\end{eqnarray}
We note that if the source obeys the covariant conservation equations
(\ref{conserve-seed1}) and (\ref{conserve-seed2}),
then equations (\ref{Einstein3}) and (\ref{Einstein4}) 
are preserved by equations (\ref{Einstein1}).

In the standard CDM model where the source is absent,
equation (\ref{Einstein1}) can be greatly simplified 
on super-horizon scales ($k\eta\ll 1$) in the radiation or matter era:
\begin{eqnarray}
  \ddot{\delta}_{\rm c}+{\frac{1}{\eta}}\dot{\delta}_{\rm c}-
  {\frac{2(2+R)}{\eta^2}}\delta_{\rm c}=0,
  & \textrm{  in radiation era,}
  \label{rad-superh}\\
  \ddot{\delta}_{\rm c}+{\frac{2}{\eta}}\dot{\delta}_{\rm c}-{\frac{6}{\eta^2}}
  \delta_{\rm c}=0,
  & \textrm{ in matter era}.
  \label{mat-superh}
\end{eqnarray}
Since $R = 3\Omega_{\rm B0} a / 4\Omega_{\rm c0} a_{\rm eq}$ by definition, 
we know $R\ll 1$ deep in the radiation era.
Thus the above equations both have a growing mode 
$\delta_{\rm c}\propto \eta^2$.
This result has an important implication for numerical simulations
of structure formation with causal sources.
In this case,
if numerical errors appear as white noise 
on super-horizon modes $k\lesssim 1/\eta$,
then they will have a growing behavior 
$S(k)=4\pi k^3 {\cal P}(k)\propto k^3\eta^4$. 
For the horizon crossing mode $k\sim 1/\eta$,
this becomes $S(k)\propto \eta$ \cite{turok}.
This means that
although energy conservation together with causality 
should forbid the growth of perturbations on super-horizon scales,
any numerical errors seeded from early times would induce
a spurious growing mode on these scales.
To overcome this problem,                       
one needs to perfectly compensate the source energy 
in the initially homogeneous background.
In the following section,
we shall discuss one of the methods that can achieve this.

\subsection{Stress-energy conservation of the pseudotensor}
\label{conservation-pseudo}

The concept of the stress-energy pseudotensor in an expanding universe
was first remarked in this context by Veeraraghavan and Stebbins \cite{VeeSte},
and further investigated by Pen, Spergel and Turok \cite{turok}.
To introduce this concept, 
we start from a perturbed Minkowski space
$\hat{g}_{\mu\nu}=\eta_{\mu\nu}+\hat{h}_{\mu\nu}$,
where the Bianchi identity $\nabla^\mu G_{\mu\nu}=0$ leads to
an ordinary conservation law $\partial^\mu G_{\mu\nu(1)}=0$ 
at linear order in $\hat{h}_{\mu\nu}$.
Adding the fact that the Einstein equations give 
$G_{\mu\nu(1)}=8\pi GT_{\mu\nu}-G_{\mu\nu(\rm nl)}$
where $G_{\mu\nu(\rm nl)}$ is the sum of non-linear terms 
in $\hat{h}_{\mu\nu}$,
we see that the right-hand side of this equation provides
an ordinarily conserved tensor, the stress-energy pseudotensor.

The generalization of this result to an FRW model is straightforward, 
with only the corrections due to the expansion of the universe.
Moving all these corrections (derivatives of the scale factor)
to the right-hand side of the Einstein equations
while keeping only the linear terms in ${h}_{\mu\nu}$,
we obtain a pseudo-stress-energy tensor 
$\tau_{\mu\nu}\equiv G_{\mu\nu(1)}/8\pi G$:
\begin{eqnarray}
  \tau_{00}
  &=&
  \frac{3}{8\pi G}\left({\frac{\dot{a}}{a}}\right)^2
  \left[\Omega_{\rm c}\delta_{\rm c}+(1+R)\Omega_{\rm r}\delta_{\rm r}\right]
  -\frac{1}{8 \pi G}\frac{\dot{a}}{a}\dot{h}+\Theta_{00} \; , 
  \label{tau00}\\
  \tau_{0i}
  &=&
  -\frac{1}{2\pi G}\left({\frac{\dot{a}}{a}}\right)^2
  (1+R)\Omega_{\rm r} v_{\rm r}^i+\Theta_{0i} \;, 
  \label{tau0i}\\
  \tau_{ij}
  &=&
  \delta_{ij}
  \frac{1}{8 \pi G}\left({\frac{\dot{a}}{a}}\right)^2
  \Omega_{\rm r}\delta_{\rm r}
  -\frac{1}{8 \pi G}
  \frac{\dot{a}}{a}(\dot{\tilde{h}}_{ij}-\frac{2}{3}\dot{h}\delta_{ij})
  +\Theta_{ij}\;.
  \label{tauij}
\end{eqnarray}
This tensor obeys an ordinary conservation law 
${\tau^{\mu\nu}}_{\!,\nu}=0$
according to the Einstein equations, or equivalently
\begin{eqnarray}
  \tau^{00}_{,0}=\tau^{0i}_{,i}\;,
  \label{conserve-tau1}\\
  \tau^{i0}_{,0}=\tau^{ij}_{,j} \;.
  \label{conserve-tau2}
\end{eqnarray}
This is not a fundamentally new conservation law,
but it describes the interchange of energy and momentum 
among the different components in the universe, 
i.e.\ 
the radiation, matter, and the source in our case.
This description appears to be physically more transparent than
the original Einstein equations.

Another advantage of invoking this formalism is that
it is easier for numerical simulations
to specify the initial conditions
and to maintain proper compensation on super-horizon scales.
As we shall explain later,
$\tau_{ij}$ can only have a white-noise power spectrum on super-horizon scales.
Thus integrating equations (\ref{conserve-tau1}) and (\ref{conserve-tau2}) 
over time shows that
$\tau_{00}$ has a $k^4$ power spectrum
and that
$\tau_{0i}$ has a $k^2$ power spectrum.
Therefore, 
as long as the horizon size at the beginning of the simulation
is smaller than the scales of our interest,
we can set $\tau_{00}=\tau_{0i}=0$ as the initial condition,
allowing for perturbations to grow only inside the horizon
and for $\tau_{00}$ to fall off as $k^4$  outside the horizon.
For simulations of structure formation with causal source,
a check of $\tau_{00}\propto k^4$ on super-horizon modes
will tell us whether or not the compensation is well obeyed.

To make use of the pseudo-stress-energy tensor formalism
in the study of cosmological perturbations,
we combine the conservation equation for radiation (\ref{entropy2}), 
the definition of pseudoenergy (\ref{tau00}),
and 
one of the alternative Einstein equations using the
pseudo-stress-energy tensor (\ref{conserve-tau1}),
to yield a convenient closed set of equations:
\begin{eqnarray}
  \ddot{s}
  & = & - \frac{\dot{R}}{1+R}\dot{s}
  +\frac{1}{3+3R} \nabla^2 (s+\delta_{\rm c})\;,
  \label{density-eqn1}\\
  \dot{\delta}_{\rm c}
  & = & 4 \pi G \frac{\dot{a}}{a}(\tau_{00}-\Theta_{00})-\frac{\dot{a}}{a}
  \left\{
  \left[\frac{3}{2}\Omega_{\rm c}+2(1+R)\Omega_{{\rm r}}\right]
  \delta_{\rm c}+
  2(1+R)\Omega_{{\rm r}}s
  \right\}, 
  \label{density-eqn2}\\
  \dot{\tau}_{00}
  & = & \Theta_{0i,i}+ \frac{1}{2\pi G}\left(\frac{\dot{a}}{a}\right)^2
  (1+R)\Omega_{{\rm r}}\dot{s}\;.
  \label{density-eqn3}
\end{eqnarray}
Here we have used equations 
(\ref{conserve-c}), (\ref{entropy}), (\ref{entropy1}) and (\ref{tau0i})
to eliminate $\dot{h}$, $\delta_{\rm r}$, $v_{\rm r}^i$ and $\tau_{0i}$ 
respectively.
By analogy to the results in Ref.~\cite{turok}, 
here we have  built both the pseudoenergy $\tau_{00}$ 
and the entropy fluctuation $s$
into the above formalism.

\subsection{Initial conditions of causal models}
\label{initial-conditions}

As required by the IUSG,
all perturbation variables are zero 
before 
any mechanism of structure formation starts to act on the
initially homogeneous and isotropic universe.
In causal models,
causality also requires that
local physical processes can never induce correlated perturbations 
on scales much larger than the horizon.
Therefore,
when the initial irregularities of the universe are first formed
(e.g.\ via the formation of cosmic defects, or the presence of inflation),
the spatial part of $\Theta_{\mu\nu}$, $\tau_{\mu\nu}$ 
and $h_{\mu\nu}$ can only have white-noise power spectra 
on super-horizon modes---their spatial perturbations being
uncorrelated on scales larger than the horizon size \cite{VeeSte}.
The same applies to $\delta_N$ and therefore $h$.
It then follows from equations (\ref{conserve-r2}), (\ref{conserve-seed2}) 
and (\ref{conserve-tau2}) respectively
that
the power spectra of ${\bf v}_{\rm r}$, $\Theta_{0i}$ and $\tau_{0i}$ 
all fall off as $k^2$ outside the horizon.
From equations (\ref{entropy1}) and (\ref{conserve-tau1}),
we also have the spectra of $s$, $\dot{s}$ and $\tau_{00}$ 
proportional to $k^4$ on these scales, as previously discussed.
As a summary, we have for super-horizon modes $k\ll 1/\eta$ that
\begin{eqnarray}
  \delta_{\rm c}, \delta_{\rm r}, h, h_{ij}, \Theta_{ij}, \tau_{ij} 
  &\propto & k^0\;,
  \\
  {\bf v}_{\rm r}, \Theta_{0i}, \tau_{0i}
  &\propto & k^1\;,
  \label{theta0i-propto}
  \\
  s, \dot{s}, \tau_{00}
  &\propto & k^2\;,
  \label{tau00-propto}
\end{eqnarray}
where `$\propto k^n$' means the power spectrum is proportional to $k^{2n}$.

Since the production time of the initial irregularities
is normally so early that
the horizon size $\eta_{\rm i}$ at that time 
is much smaller than the cosmological scales $k_{\rm cos}^{-1}$ 
of our interest (i.e.\ $k_{\rm cos} \eta_{\rm i} \ll 1$),
the above conditions can be regarded as general initial conditions
for all scales of cosmological interest.
If we require $k_{\rm cos}\eta_{\rm i} \ll 1$ in our analysis,
we can simply choose
\begin{equation}
  \label{initial-condition}
  v_{\rm r}^i = \Theta_{0i} = \tau_{0i} = s =  \dot{s} = \tau_{00} = 0,
\end{equation}
as the initial conditions,
because their power spectra all decay as either $k^2$ or $k^4$
outside the horizon.

With such a choice,
we can see from equation (\ref{tau00}) that
there is still freedom for the choice of 
$\delta_{\rm c}$, $\delta_{\rm r}$ and $\dot{h}$ 
into which to compensate $\Theta_{00}$
when $\Theta_{00}$ was first formed.
Nevertheless,
as we shall analytically prove later,
no matter how $\Theta_{00}$ was compensated into 
the background contents of the universe
when the causal source was first formed,
the resulting matter density perturbations today would be the same.
We note that 
this was first numerically observed in Ref.~\cite{turok},
and here we shall provide a thorough interpretation to it
using our analytical solutions to be obtained later.
We also note that none of the above arguments will hold 
if the initial perturbations are seeded in an acausal way,
which is nevertheless not of our current interest.

\subsection{approximation of instantaneous decoupling}
\label{instantaneous-decoupling}

One thing we have not included in our formalism is the treatment
at and after the decoupling epoch $\eta_{\rm d}$.
Before this epoch,
photons and baryons are assumed to be tightly coupled,
forming a single $\gamma$B fluid.
At the decoupling epoch $\eta_{\rm d}$,
baryons and photons are assumed to be
instantaneously decoupled from each other,
so that $\delta_\gamma$ and $\delta_{\rm B}$ evolve separately afterwards.
A numerical fit to the redshift of the decoupling epoch is \cite{EisensteinHu}
\begin{eqnarray}
  z_{\rm d}
  & = & 1291{\frac{(\Omega_{\rm m0}h^2)^{0.251}}{1+0.659(\Omega_{\rm m0}
      h^2)^{0.828}}} \left[1+b_1(\Omega_{\rm B0}h^2)^{b_2}\right],
  \label{z-d}\\
  b_1 & = & 0.313(\Omega_{\rm m0}h^2)^{-0.419}
  \left[1+0.607(\Omega_{\rm m0}h^2)^{0.674}\right],\\
  b_2 & = & 0.238(\Omega_{\rm m0}h^2)^{0.223}.
\end{eqnarray}
Although this is the result for the decoupling epoch of baryons
and there is another fit for that of photons,
these two epochs---the recombination of baryons and the last scattering of photons---coincide
approximately in the absence of subsequent reionization 
\cite{HuSugiyama1,HuSugiyama2}.

In addition,
the photons and baryons are not in fact perfectly coupled, and 
this leads to the diffusion damping of photons and Silk damping of baryons
\cite{Silk} during the decoupling epoch.
To model these effects,
we apply damping envelopes to both $\delta_\gamma$ and $\delta_{\rm B}$
at the decoupling epoch $z_{\rm d}$, i.e.\
\begin{equation}
  \label{damping-effects}
  \hat\delta_{N \rm (d)}=\widetilde{\delta}_{N\rm (d)} {\cal D}_N(k),
  \quad N=\gamma, \; {\rm B},
\end{equation}
where the tilde indicates the Fourier transform of a quantity
and $k$ is the wave number.
The photon diffusion damping envelope can be approximated by the form
\cite{HuSugiyama2}
\begin{equation}
  \label{D_gamma}
  {\cal D}_{\gamma}(k)\simeq e^{-(k/k_{\gamma})^{m\gamma}},
\end{equation}
where
\begin{eqnarray}
  \frac{k_{\gamma}}{{\rm Mpc}^{-1}} & = & \left\{ {\frac{2}{\pi}}
    \arctan\left[
      {\frac{\pi}{2}}\left({\frac{F_2}{F_1}}\right)^{p2/p1}
      (\Omega_{\rm B0}h^2)^{p2}\right] \right\}^{p1/p2}F_1\;,\\
  m_{\gamma} & = & 1.46(\Omega_{\rm m0}h^2)^{0.0303}
  \left(1+0.128\arctan\left\{\ln\left[(32.8\Omega_{\rm B0}
        h^2)^{-0.643}\right]\right\}\right),\\
  p_1 & = & 0.29,\\
  p_2 & = & 2.38 (\Omega_{\rm m0}h^2)^{0.184},\\
  F_1 & = & 0.293(\Omega_{\rm m0}h^2)^{0.545}
  \left[1+(25.1\Omega_{\rm m0}h^2)^{-0.648}\right],\\
  F_2 & = & 0.524(\Omega_{\rm m0}h^2)^{0.505}
  \left[1+(10.5\Omega_{\rm m0}h^2)^{-0.564}\right].
\end{eqnarray}
Silk damping for the baryons can likewise be approximated as
\cite{HuSugiyama2}
\begin{equation}  
  \label{D_B}
  {\cal D}_{\rm B}(k)\simeq e^{-(k/k_{\rm S})^{m_{\rm S}}},
\end{equation}
where
\begin{eqnarray}
  \frac{k_{\rm S}}{\rm Mpc^{-1}} & = &
  1.38(\Omega_{\rm m0}h^2)^{0.398}(\Omega_{\rm B0}h^2)^{0.487}
  \frac{1+(96.2\Omega_{\rm m0}h^2)^{-0.684}}
  {1+(346\Omega_{\rm B0}h^2)^{-0.842}},\\
  m_{\rm S} & = & 1.40\frac{(\Omega_{\rm B0}h^2)^{-0.0297}
    (\Omega_{\rm m0}h^2)^{0.0282}}{1+
    (781\Omega_{\rm B0}h^2)^{-0.926}}.
\end{eqnarray}

In some scenarios with causal sources,
the damping envelopes (\ref{D_gamma}) and (\ref{D_B}) 
may depart from the form of exponential fall-off here
to a power-law decay towards smaller scales.
This is due to the survival of perturbations
which are actively seeded during the decoupling process.
For example,
in models with cosmic strings,
the departure appears on scales smaller than of order a few arc-minutes
(i.e.\ the multipole index $l\gtrsim 3000$)
\cite{damping-power-law}.
Certainly this is beyond the scale range of our interest.
Moreover,
since the decoupling process is relatively a short instant
in the entire evolution history of the perturbations,
the contribution from these survived small-scale perturbations
should be relatively small.
Adding the fact that
we expect the post-decoupling contribution
in the perturbations seeded by defects
to have a power-law fall-off on small scales
due to a certain topology of the source \cite{AveShe3},
the small-scale power in the final perturbations
is likely to be dominated by this post-decoupling contribution,
rather than
the primary perturbations 
(those seeded before and during the decoupling,
whose power spectrum
first exponentially decays and then turns to a power-law fall-off).
Therefore,
on the scales of our interest,
the damping approximation employed here
should be still appropriate for models with cosmic defects.

Now we consider the evolution of $\delta_\gamma$ and $\delta_{\rm B}$
after the decoupling epoch $z_{\rm d}$.
From the energy conservation law (\ref{conserve-1})--(\ref{conserve-3}),
we have for the baryon perturbations
\begin{equation}
  \ddot \delta_{\rm B} - \ddot \delta_{\rm c}
   +  {\dot{a} \over a} (\dot \delta_{\rm B}-\dot \delta_{\rm c})
   =0.
  \label{conserve-B}
\end{equation}
This implies $(\dot\delta_{\rm B}- \dot \delta_{\rm c})\propto a^{-1}$,
meaning that
the evolution of $\delta_{\rm B}$ and $\delta_{\rm c}$ will soon converge to
the same behavior.
We also know that matter perturbations grow as $\eta^{2}$ in the matter era
so that $[\delta_{\rm B(d)}- \delta_{\rm c(d)}]$ is relatively small
when compared with either $\delta_{\rm B0}$ or $\delta_{\rm c0}$.
As a consequence,
in the calculation of $\delta_{\rm B0}$ and $\delta_{\rm c0}$ to linear order,
it is appropriate
to combine
$\delta_{\rm B}$ and $\delta_{\rm c}$
at the decoupling epoch $z_{\rm d}$ as
\begin{equation}
  \label{delta-m-decoupling}
  \widetilde\delta_{\rm m(d)}
  =
  \frac{\Omega_{\rm B0}\hat\delta_{\rm B(d)}
    +\Omega_{\rm c0}\widetilde\delta_{\rm c(d)}}
  {\Omega_{\rm B0}+\Omega_{\rm c0}}
  =
   \frac{3\Omega_{\rm B0}\widetilde\delta_{\gamma \rm(d)}{\cal D}_{\rm B}/4
     +\Omega_{\rm c0}\widetilde\delta_{\rm c(d)}}
  {\Omega_{\rm B0}+\Omega_{\rm c0}},
\end{equation}
and the same for their time derivatives.
Then we have only two fluids after the decoupling:
the photon fluid ($\Omega_\gamma$) and
the matter fluid,
which is linearly combined from the CDM and baryon fluids
($\Omega_{\rm m}=\Omega_{\rm c}+\Omega_{\rm B}$).
Eventually
we can take the matter perturbations at the present epoch to be
$\widetilde\delta_{\rm c0}
\equiv \widetilde\delta_{\rm B0}
\equiv \widetilde\delta_{\rm m0}$.

To sum up,
we first evolve the CDM and $\gamma$B perturbations
up to the decoupling epoch $z_{\rm d}$ given by (\ref{z-d}),
noting that our formalism extracts the photon component $\gamma$
from the $\gamma$B fluid.
We then apply damping envelopes to
$\widetilde\delta_{\gamma\rm(d)}$ and $\widetilde\delta_{\rm c(d)}$,
as illustrated by equation (\ref{damping-effects}),
to account for the photon diffusion and Silk damping.
$\widetilde\delta_{\rm m(d)}$ is then obtained by 
linearly combining 
$\widetilde\delta_{\rm c(d)}$ and $\widetilde\delta_{\rm B(d)}$,
as shown in equation (\ref{delta-m-decoupling}).
Finally we carry on the evolution of $\widetilde\delta_{\rm m}$
and $\widetilde\delta_{\rm r}$ from the epoch $z_{\rm d}$ to the present,
using our previous perturbation equations with $R=0$ and
the subscript `c' replaced by `m'.

\subsection{Accuracy for the standard CDM models}
\label{verification-stdcdm}

\begin{figure}
  \centering\epsfig{figure=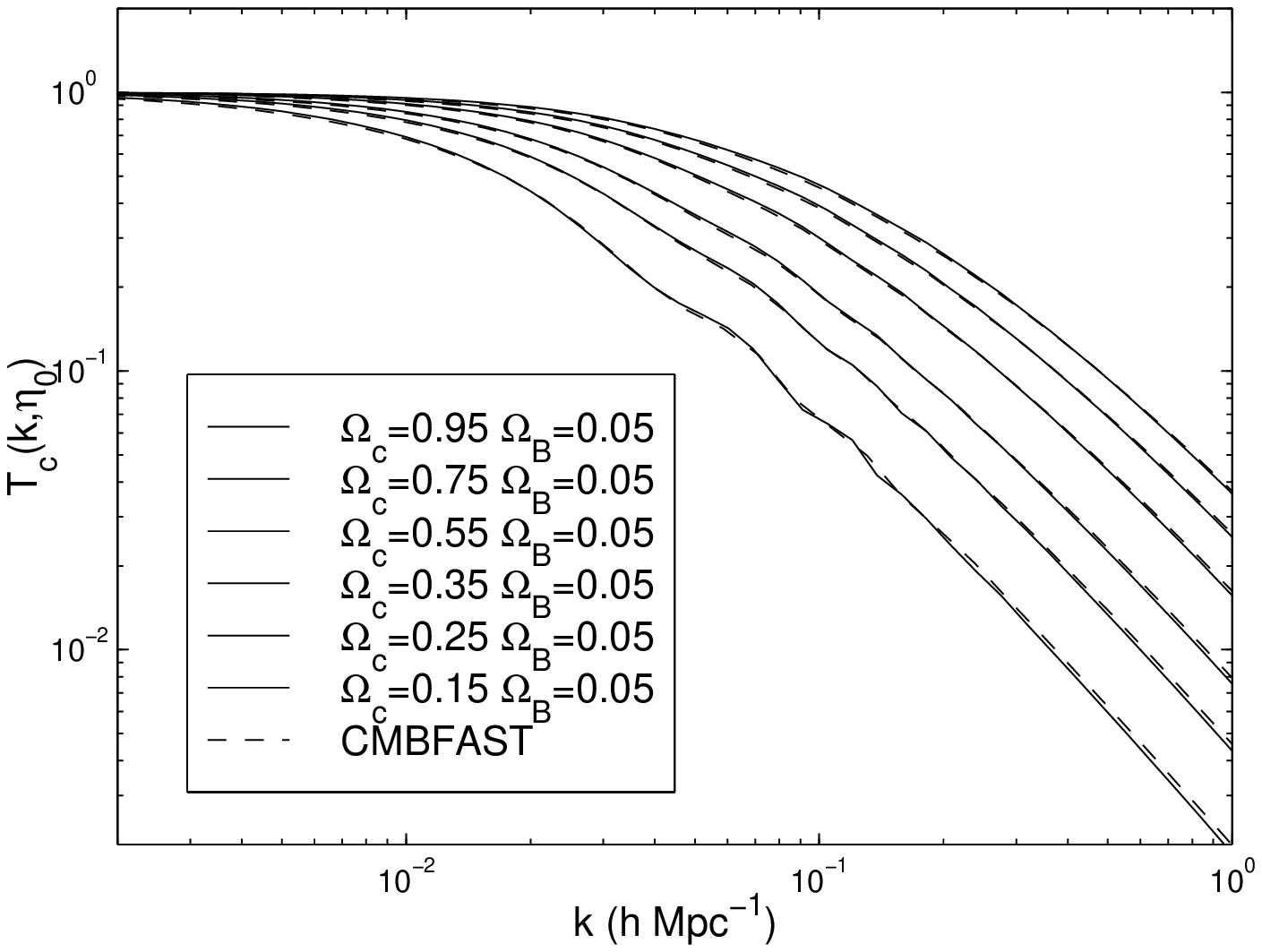, width=120mm}\\
  \centering\epsfig{figure=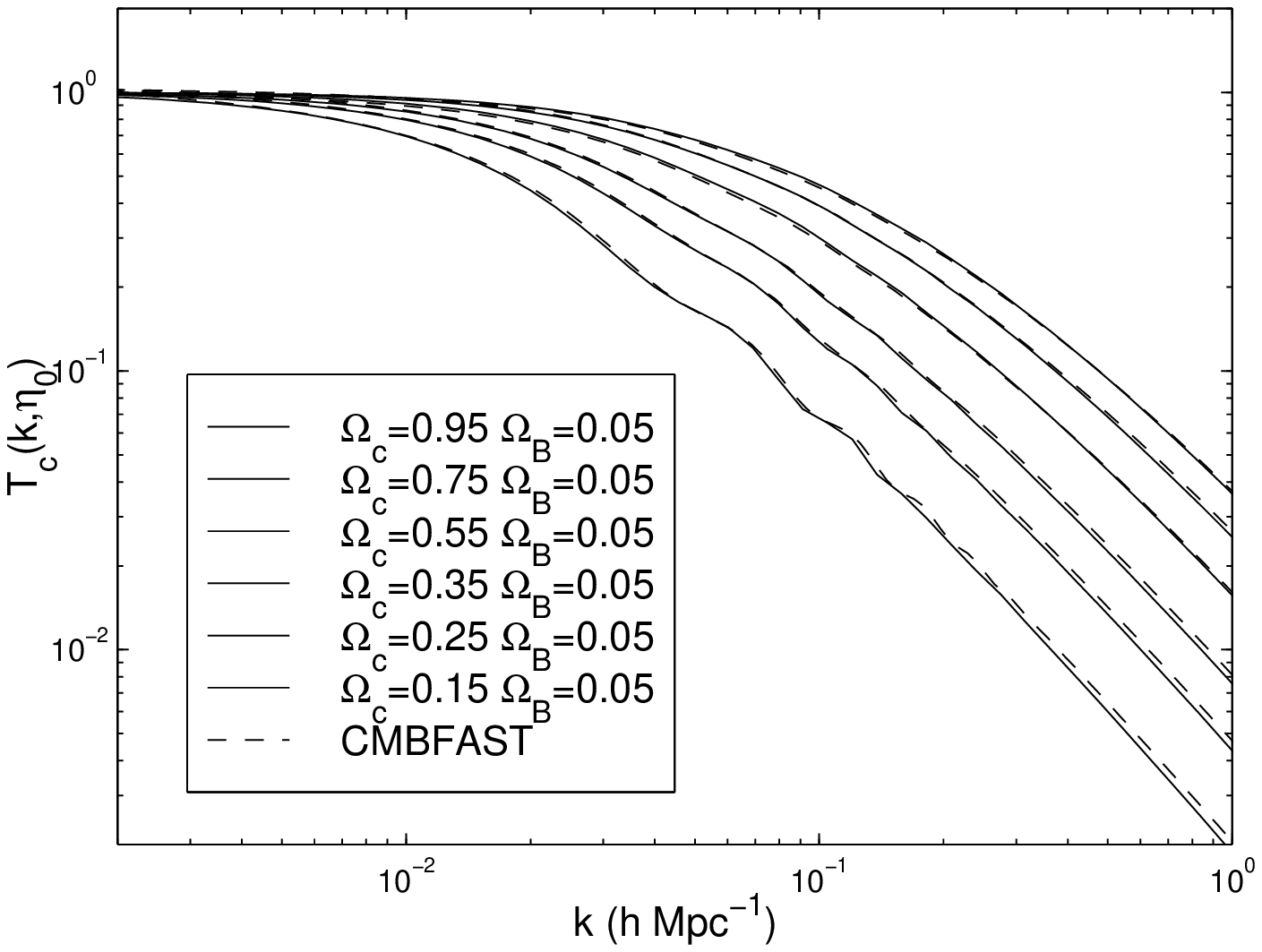, width=120mm}
  \caption[]
  {Comparison of our CDM transfer functions at the present epoch
    $T_{\rm c}(k,\eta_0)$
    with results obtained from CMBFAST \cite{cmbfast}.
    On the top are results in flat models with a cosmological constant
    (i.e.\ $\Omega_{\Lambda 0}+\Omega_{\rm c0}+\Omega_{\rm B0}=1$).
    At the bottom are results in open models without a cosmological constant.
    Results using our formalism are plotted as solid lines,
    while the results from CMBFAST are plotted as dashed lines.
    We have used $h=0.7$ throughout.
    The mass fraction of Helium-4 $Y_{\rm He}=0.24$ and
    the number of neutrino species $N_\nu=3.04$ have been used
    in obtaining the results from CMBFAST.
    }
  \label{fig-transTc}
\end{figure}

To verify our scheme for evolving cosmological perturbations,
we first calculate the CDM transfer function
in the context of the adiabatic inflationary CDM model:
\begin{equation}
  \label{cdm-transfer-function}
    T_{\rm c}(k,\eta_0)=
    \frac{\widetilde{\delta}_{\rm c}(k, \eta_0)
      \widetilde{\delta}_{\rm c} (0, 0)}
    {\widetilde{\delta}_{\rm c} (k, 0)
      \widetilde{\delta}_{\rm c} (0, \eta_0)},
\end{equation}
where $\eta_0$ is the present conformal time.
To this end, we employ equations
(\ref{density-eqn1}), (\ref{density-eqn2}) and (\ref{density-eqn3})
in the absence of the source terms,
and the approximation of instantaneous decoupling described above.
We start the evolution in the deep radiation era
when $\Omega_{\rm m} \ll \Omega_{\rm r}\approx 1$, $R \ll 1$,
and $\eta_{\rm i}\ll 1/k$ for a given mode $k$.
In this case,
one choice of the initial conditions is
\begin{equation}
  \label{cdm-initial-conditions}
  s=\dot s=0, \quad
  \delta_{\rm c}=\eta_{\rm i}^2, \quad
  \tau_{00}=\frac{1}{\pi G}.
\end{equation}
Figure~\ref{fig-transTc} shows
our results for the CDM transfer functions $T_{\rm c}(k,\eta_0)$
at the present epoch in different cosmologies,
together with the results obtained from CMBFAST \cite{cmbfast}.
It is clear that they agree very well.
The discrepancy of the two reaches its maximum of about 5\%
at the scale $k\approx 1 h$Mpc$^{-1}$
in the open model with $\Omega_{\rm c0}=0.15$ and $\Omega_{\rm B0}=0.05$.
We have also checked our results against those in Ref.~\cite{BonEfs},
and they are in agreement again within a $5$\% error.
In addition,
from the bottom curves in Figure~\ref{fig-transTc},
we notice the oscillations resulting from the photon-baryon coupling
before $\eta_{\rm d}$ in cosmologies with high baryon fractions
$\Omega_{\rm B0}/\Omega_{\rm m0}$.

\begin{figure}
  \centering\epsfig{figure=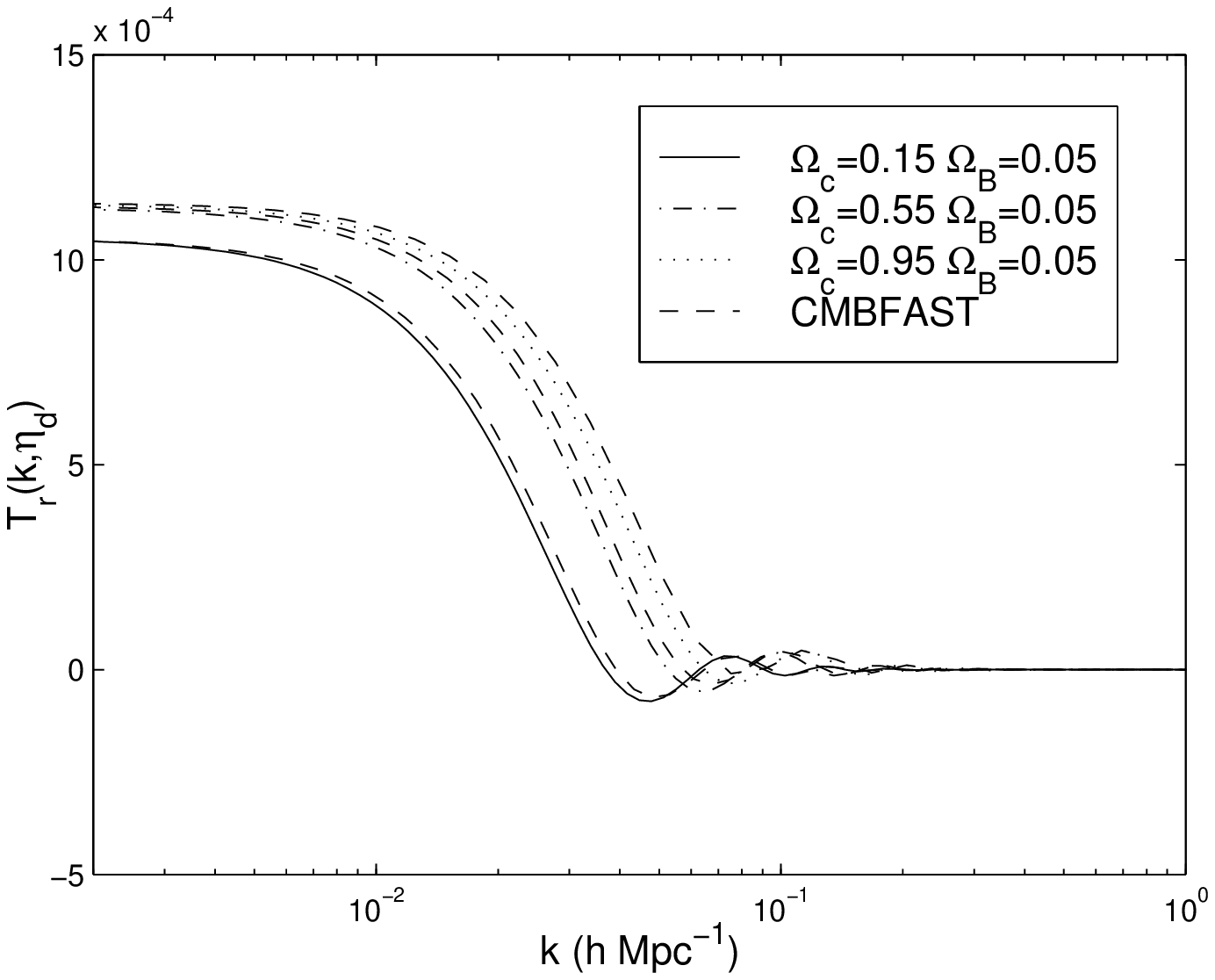, width=120mm}\\
  \centering\epsfig{figure=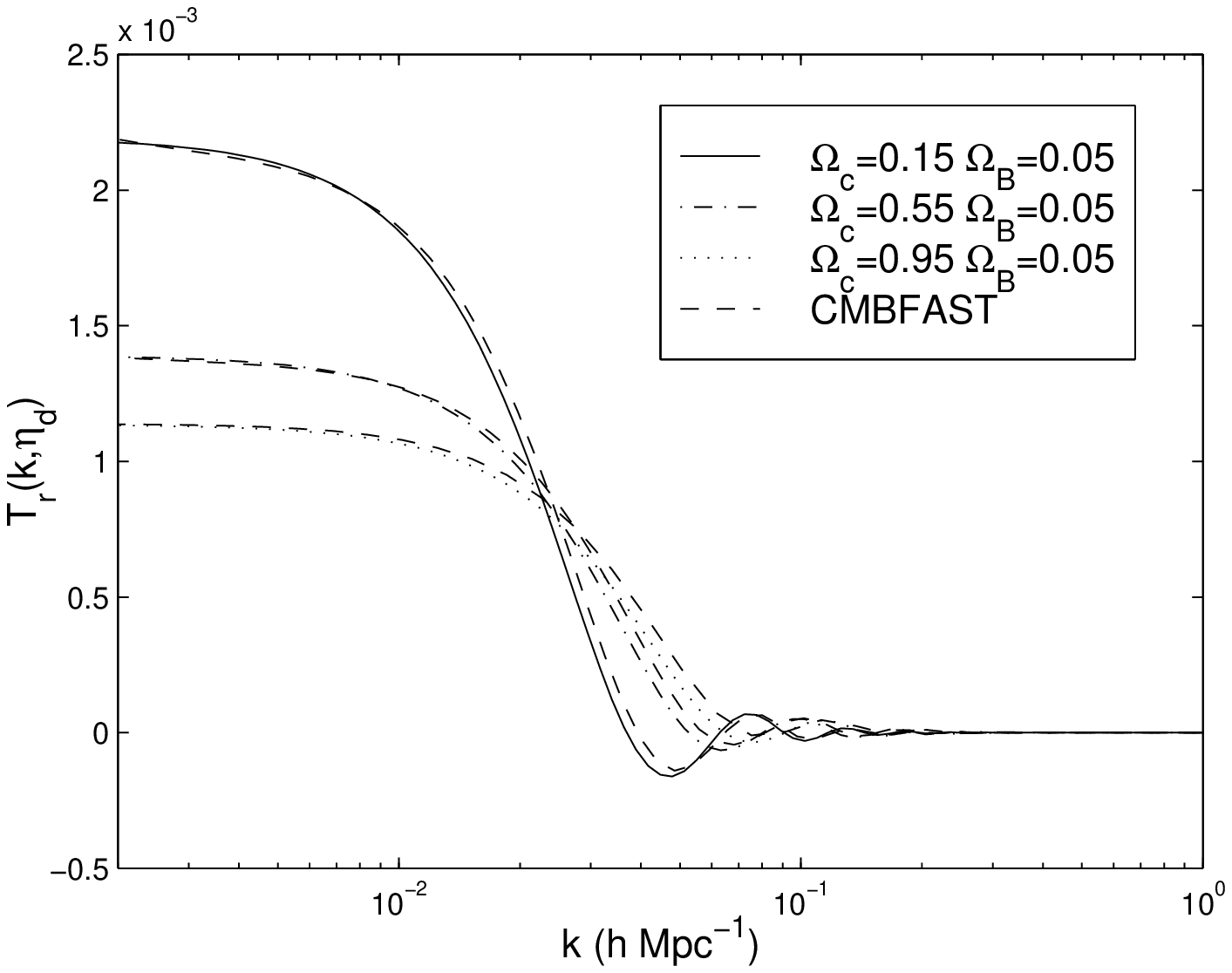, width=120mm}
  \caption[]
  {Comparison of our radiation transfer functions at the decoupling epoch
    $T_{\rm r}(k,\eta_{\rm d})$
    with results obtained from CMBFAST \cite{cmbfast}.
    On the top are results in flat models with a cosmological constant
    (i.e.\ $\Omega_{\Lambda 0}+\Omega_{\rm c0}+\Omega_{\rm B0}=1$).
    At the bottom are results in open models without a cosmological constant.
    Our results are plotted as solid, dot-dashed, and dotted lines,
    while the CMBFAST results are plotted as dashed lines.
    Note that this transfer function has been normalised
    to both the amplitude of  the super-horizon CDM perturbations today
    and the initial CDM power spectrum
    (see eq.~[\ref{rad-transfer-function}]).
    }
  \label{fig-transTr}
\end{figure}

Next, we calculate the radiation transfer function at the decoupling epoch,
since the radiation perturbations at this epoch will appear as the intrinsic
CMB anisotropies.
We define this transfer function as:
\begin{equation}
  \label{rad-transfer-function}
    T_{\rm r}(k,\eta_{\rm d})=
    \frac{\widetilde{\delta}_{\rm r}(k, \eta_{\rm d})
      \widetilde{\delta}_{\rm c} (0, 0)}
    {\widetilde{\delta}_{\rm c} (k, 0)
      \widetilde{\delta}_{\rm c} (0, \eta_0)},
\end{equation}
where we normalize the radiation perturbations at $\eta_{\rm d}$
to both the amplitude of the super-horizon CDM perturbations today
and the initial CDM power spectrum,
as we did for $T_{\rm c}(k,\eta_0)$ (see eq.~[\ref{cdm-transfer-function}]).
This definition will enable us to verify not only the scale dependence
of the evolution of perturbations,
but also their normalizations.
Figure~\ref{fig-transTr} shows our results,
again as a comparison with the results from CMBFAST.
We see that
although the scale dependence of our results
is slightly different from that of the CMBFAST results,
the overall normalisation appears to be quite accurate.
The sideway shift of the oscillatory peaks in our results
when compared with the peaks from CMBFAST
has a maximum of about 5\%
in the flat model with $\Omega_{\rm c0}=0.95$ and $\Omega_{\rm B0}=0.05$.
This discrepancy results naturally from
the instantaneous-decoupling approximation in our formalism.
As a result,
despite the small inaccuracy,
our formalism provides a much more numerically efficient way
than the full Einstein-Boltzmann scheme
in calculating the density perturbations.

\section{Solutions of matter perturbations}
\label{solutions-of-matter-perturbations}
\subsection{Decomposition of perturbations}
\label{evolution-equations-and-decomposition-of-perturbation}

We first consider density perturbations about a flat FRW model
with a cosmological constant $\Lambda$,
which are causally sourced by an evolving source field
with the energy-momentum tensor $\Theta_{\mu\nu}({\bf x}, \eta)$.
As seen in the previous section,
with the photon-baryon tight coupling approximation
in the synchronous gauge,
the linear evolution equations of the radiation and CDM perturbations
can be given by equations
(\ref{density-eqn1}), (\ref{density-eqn2}) and (\ref{density-eqn3}),
which are derived from equations (\ref{entropy2}), (\ref{tau00})
and (\ref{conserve-tau1}).
This set of equations has the advantage
in controling the initial condition for numerical simulations,
as well as understanding the law of stress-energy conservation.
For analytic simplicity, however,
we shall drop the use of $\tau_{00}$ in this section,
and employ equations (\ref{conserve-r}) and (\ref{Einstein1})
to form an alternative set of evolution equations for density perturbations:
\begin{eqnarray}
  \ddot \delta_{\rm r} - {4 \over 3} \ddot \delta_{\rm c}
  +\frac{\dot R}{1+R}(\dot \delta_{\rm r} - {4 \over 3} \dot \delta_{\rm c})
  - {1 \over 3(1+R)} \nabla^2 \delta_{\rm r}
  =  0,
  \label{delta-one}\\
  \ddot \delta_{\rm c}
  + {\dot a \over a} \dot \delta_{\rm c} 
  - {3 \over 2}\Big({\dot a \over a}\Big)^2 \,
  \left[
    \Omega_{\rm c}\delta_{\rm c} + (2+R)\Omega_{\rm r}\delta_{\rm r}
  \right]
   =  4 \pi G \Theta_+.
  \label{delta-two}
\end{eqnarray}
We note again that the cosmological constant $\Lambda$
affects only the background dynamics 
(i.e., the evolution of the scale factor $a$),
but does not contribute extra terms in the above perturbation equations.
After the decoupling epoch $\eta_{\rm d}$,
the treatment is essentially the same as that introduced
in section~\ref{instantaneous-decoupling}.
We have numerical verified 
in the context of the adiabatic inflationary CDM model that
the set of equations (\ref{delta-one}) and (\ref{delta-two})
and the set of equations 
(\ref{density-eqn1}), (\ref{density-eqn2}) and (\ref{density-eqn3})
indeed give identical
transfer functions of density perturbations,
with a numerical discrepancy of less than $0.1$\%.

Assuming that
the causal source was formed at some initial time $\eta_{\rm i}$
and then evolved to the current time $\eta$,
it proves useful to split the source-seeded linear perturbations into
initial (I) and subsequent (S)
parts \cite{VeeSte}:
\begin{equation}
  \delta_N ({\bf x}, \eta) = \delta_N^{\rm I}({\bf x}, \eta)
  + \delta_N^{\rm S}({\bf x}, \eta)\,, \; N={\rm c, r}.
  \label{deltaN}
\end{equation}
The initial perturbations $\delta_N^{\rm I}({\bf x}, \eta)$
originate from the source
configuration at $\eta_{\rm i}$,
while the subsequent perturbations $\delta_N^{\rm S}({\bf x}, \eta)$
are actively and cumulatively seeded by
the later evolution of the source at each $\hat{\eta}$,
where $\eta_{\rm i}<\hat{\eta}<\eta$.
This is equivalent to having the initial conditions
\begin{eqnarray}
  \delta_N^{\rm I}(\eta_{\rm i})=\delta_N(\eta_{\rm i}),
  \quad
  \dot{\delta}_N^{\rm I}(\eta_{\rm i})=\dot{\delta}_N(\eta_{\rm i}),
  \label{inic_deltaI}
  \\
  \delta_N^{\rm S}(\eta_{\rm i})=\dot{\delta}_N^{\rm S}(\eta_{\rm i})=0.
  \label{inic_deltaS}
\end{eqnarray}
Because the source induces isocurvature perturbations,
$\delta^{\rm I}({\bf x}, \eta)$
must compensate  $\delta^{\rm S}({\bf x}, \eta)$ on comoving scales
$|{\bf x}-{\bf x}'|>\eta$ to prevent acausal perturbation growth on
super-horizon scales.
One of the aims of this paper is to show analytically
how this compensation mechanism can be achieved.
Now 
we can solve the system of equations (\ref{delta-one}) and (\ref{delta-two})
by employing the integral equation with Green functions:
\begin{eqnarray}
  \delta^{\rm I}_{N}({\bf x},\eta) & = &
  \sum_{N'}
  \int \!\!d^3x'\: {\cal G}^{NN'}_1 (X;\eta,\eta_{\rm i})
  \delta_{N'}({\bf x'},\eta_{\rm i})
  +
  \int \!\!d^3x'\: {\cal G}^{NN'}_2 (X;\eta,\eta_{\rm i})
  \dot\delta_{N'}({\bf x'},\eta_{\rm i}),
  \label{delta_I_N0} \\ 
  \delta^{\rm S}_{N}({\bf x},\eta) & = &
  4 \pi G \int_{\eta_{\rm i}}^{\eta} d\hat{\eta}
  \int d^3x'\: {\cal G}^{N{\rm s}} (X;\eta,\hat{\eta})
  \Theta_{+}({\bf x'},\hat{\eta}),
  \label{delta_S_N0}
\end{eqnarray}
where $X=|{\bf x}-{\bf x'}|$.
The easiest method of obtaining the Green-function solutions is to go to
Fourier space and solve the resulting homogeneous system of
ordinary differential equations
with appropriate initial conditions.
Since the Green functions depend only on
the modulus of $X=|{\bf x}-{\bf x'}|$, it follows that their Fourier
amplitudes must depend only on the modulus of ${\bf k}$.
Thus we have
\begin{eqnarray}
  \widetilde{\delta}^{\rm I}_{N}({\bf k},\eta)
  & = &
  \sum_{N'}
  \left[
    \widetilde{\cal G}^{NN'}_1 (k;\eta,\eta_{\rm i})
    \widetilde{\delta}_{N'}({\bf k},\eta_{\rm i})
    +
    \widetilde{\cal G}^{NN'}_2 (k;\eta,\eta_{\rm i})
    \dot{\widetilde{\delta}}_{N'}({\bf k},\eta_{\rm i})
  \right]
  ,
  \label{delta_I_N} \\ 
  \widetilde{\delta}^{\rm S}_{N}({\bf k},\eta)
  & = &
  4 \pi G \int_{\eta_{\rm i}}^{\eta} 
  \widetilde{\cal G}^{N{\rm s}} (k;\eta,\hat{\eta})
  \widetilde\Theta_{+}({\bf k},\hat{\eta})
  \, d\hat{\eta}\,.
  \label{delta_S_N}
\end{eqnarray}
We notice that equation (\ref{delta_S_N}) is different from
the form in Ref.~\cite{VeeSte},
where the authors identified our $\widetilde{\cal G}^{N{\rm s}}$
as $\widetilde{\cal G}_2^{N{\rm c}}$.
This identification is incorrect,
because 
$\widetilde{\cal G}^{N{\rm s}}$ and $\widetilde{\cal G}_2^{N{\rm c}}$
have different initial conditions, as we shall see.

For simplicity,
we assume no baryons and therefore set $R=0$ for now,
and shall relax this constraint later.
With the change of variable $y=1+A\eta/2$
where $A=2(\sqrt{2}-1)/\eta_{\rm eq}$ (leading to $a/a_{\rm eq}=y^2-1$),
and with the formalism (\ref{delta_I_N}) and (\ref{delta_S_N}),
we can rewrite 
equations (\ref{delta-one}) and (\ref{delta-two})
in Fourier space as
\begin{eqnarray}
  {\widetilde{\cal G}^{\rm r}}{''}
  - {4\over 3}{\widetilde{\cal G}^{\rm c}}{''}
  + {4 k^2 \over 3A^2} \widetilde{\cal G}^{\rm r} = 0\ ,
  \label{G_two}\\
  (1-y^2){\widetilde{\cal G}}^{\rm c}{''}-2y{\widetilde{\cal G}}^{\rm c}{'}+
  \left[ 6 -
    \frac{12{\widetilde{\cal G}}^{\rm r}/{\widetilde{\cal G}}^{\rm c}}{1-y^2}
  \right]{\widetilde{\cal G}}^{\rm c} = 0\ ,
  \label{G_one}
\end{eqnarray}
where a prime represents a derivative with respect to $y$,
${\widetilde{\cal G}}^{\rm c}\equiv
{\widetilde{\cal G}}^{{\rm c}N}_1$,
${\widetilde{\cal G}}^{{\rm c}N}_2$
or
${\widetilde{\cal G}}^{{\rm cs}}$,
and
${\widetilde{\cal G}}^{\rm r}\equiv
{\widetilde{\cal G}}^{{\rm r}N}_1$,
${\widetilde{\cal G}}^{{\rm r}N}_2$
or
${\widetilde{\cal G}}^{{\rm rs}}$.
According to equations (\ref{delta_I_N}) and (\ref{delta_S_N}),
the initial conditions (\ref{inic_deltaI}) and (\ref{inic_deltaS})
now become:
\begin{eqnarray}
  {\widetilde{\cal G}}^{\rm cc}_1=
  \dot{\widetilde{\cal G}^{\rm cc}_2}=
  {\widetilde{\cal G}}^{\rm rr}_1=
  \dot{\widetilde{\cal G}^{\rm rr}_2}=
  1
  \quad & {\rm at} &  \quad \eta=\eta_{\rm i},
  \label{GNNini}\\
  \dot{\widetilde{\cal G}^{\rm cs}}=
  \frac{3}{4}\dot{\widetilde{\cal G}^{\rm rs}}=
  1
  \quad & {\rm at} &  \quad \eta=\hat{\eta},
  \label{GNsini}
\end{eqnarray}
with all the other Green functions and their time derivatives vanishing.
There are three things we should notice here.
First, it is required 
that $\widetilde{\cal G}^{NN'}_i(k;\eta,\eta_{\rm i})=0$
for $\eta \leq \eta_{\rm i}$,
and 
that $\widetilde{\cal G}^{N{\rm s}}(k;\eta,\hat{\eta})=0$
for $\eta \leq \hat{\eta}$.
Second,
the Green functions $\widetilde{\cal G}^{NN'}_i$ 
only describe the time dependence of the homogeneous version of
equations (\ref{delta-one}) and (\ref{delta-two}),
while the Green functions $\widetilde{\cal G}^{N{\rm s}}$
are, by the conventional definition, the true Green functions
used to solve the inhomogeneous equations 
(\ref{delta-one}) and (\ref{delta-two}).
Finally,
since there are only four variables 
in equations (\ref{G_two}) and (\ref{G_one}) 
(i.e.\ 
${\widetilde{\cal G}}^{\rm c}$,
${\widetilde{\cal G}}^{\rm r}$,
$\dot{\widetilde{\cal G}}^{\rm c}$ and
$\dot{\widetilde{\cal G}}^{\rm r}$),
there must exist some dependence among the five sets of Green functions
(i.e.\ 
${\widetilde{\cal G}}^{N\rm c}_1$,
${\widetilde{\cal G}}^{N\rm r}_1$,
${\widetilde{\cal G}}^{N\rm c}_2$,
${\widetilde{\cal G}}^{N\rm r}_2$ and
${\widetilde{\cal G}}^{N\rm s}$).
This dependence can be observed from the initial conditions 
(\ref{GNNini}) and (\ref{GNsini}), 
which yield
\begin{equation}
  \label{Gcs-2}
  \widetilde{\cal G}^{N \rm s}
  =
  \widetilde{\cal G}^{N \rm c}_2
  +\frac{4}{3}\widetilde{\cal G}^{N\rm r}_2\,.
\end{equation}
In Ref.~\cite{VeeSte},
the authors ignored the fact that
$\dot{\widetilde{\cal G}^{\rm rs}}=4/3$
in the initial condition (\ref{GNsini}).
This ignorance led to
the absence of the second term in equation (\ref{Gcs-2})
(and thus the identification of 
$\widetilde{\cal G}^{N{\rm s}}=\widetilde{\cal G}_2^{N{\rm c}}$),
and consequently
the incorrect solutions of Green functions in their final results.
Based on equations (\ref{G_two}) and (\ref{G_one})
with the initial conditions (\ref{GNNini}) and (\ref{GNsini}),
in the following subsections
we shall analytically derive a complete set of  Green-function solutions 
for the matter perturbations,
which will then be numerically verified.

\subsection{Super-horizon and sub-horizon modes}
\label{super-horizon-and-sub-horizon-modes}

Under the limit $k\eta\gg 1$ or $k\eta \ll 1$,
the ratio ${\widetilde{\cal G}}^{\rm r}/{\widetilde{\cal G}}^{\rm c}$
will approach a constant (see below),
so that equation (\ref{G_one}) becomes the associated Legendre equation,
with solutions composed of the associated Legendre functions
$P_2^{-\mu}(y)$ and $Q_2^{\mu}(y)$, where
$\mu=\sqrt{12{\widetilde{\cal G}}^{\rm r}/{\widetilde{\cal G}}^{\rm c}}$.
We shall use subscripts $\infty$ and $0$ to denote solutions in the
limits $k\eta\gg 1$ and $k\eta \ll 1$ respectively.
For simplicity,
we shall denote both $\hat{\eta}$ and $\eta_{\rm i}$ as $\hat{\eta}$
in the following solutions.
\begin{enumerate}
\item $k\eta\gg 1$:
  When the wavelengths are much smaller than the horizon size,
  the radiation oscillates many times per expansion time and
  its effect is therefore negligible.
  By setting ${\widetilde{\cal G}}^{\rm r}/{\widetilde{\cal G}}^{\rm c}=0$,
  equation (\ref{G_one}) can be solved as
  \begin{equation}
    \label{GcNinf}
    {\widetilde{\cal G}}^{\rm c}_{\infty}(\eta,\hat{\eta})=
    E(\hat{\eta})P^0_2(y)+
    F(\hat{\eta})Q^0_2(y),
  \end{equation}
  where $E(\hat{\eta})$ and $F(\hat{\eta})$
  are functions of $\hat{\eta}$.
  This gives the sub-horizon solutions.
\item $k\eta\ll 1$:
  When the wavelengths are much longer than the horizon size,
  we have ${\widetilde{\cal G}}^{\rm r}/{\widetilde{\cal G}}^{\rm c}=4/3$
  as the consequence of zero entropy (see eqs.~[\ref{entropy}] and [\ref{initial-condition}]), giving $\mu=4$.
  Thus equations (\ref{G_two}) and (\ref{G_one}) yield
  \begin{eqnarray}
    {\widetilde{\cal G}}^{\rm r}_{0}(\eta,\hat{\eta}) & = &
    {4 \over 3}{\widetilde{\cal G}}^{\rm c}_{0}(\eta,\hat{\eta})
    + \alpha_{\rm i}(\eta-\hat{\eta})
    + \beta_{\rm i}\;,
    \label{Gr0}\\
    {\widetilde{\cal G}}^{\rm c}_{0}(\eta,\hat{\eta}) & = &
    G(\hat{\eta})P^{-4}_2(y)+
    H(\hat{\eta})Q^4_2(y)
    \nonumber\\
      +&
      \!\!12&
      \!\!\!\!\int_{y_{\rm i}}^{y}
    \frac{Q^4_2(x)P^{-4}_2(y)-P^{-4}_2(x)Q^4_2(y)}
    {Q^4_2(x)P^{-4'}_2(x)-P^{-4}_2(x)Q^{4'}_2(x)}
    \frac{A\beta_{\rm i}+2\alpha_{\rm i}(x-y_{\rm i})}
    {A(x^2-1)^2}
    dx,
    \label{Gc0}
  \end{eqnarray}
  where $\alpha_{\rm i}$ and $\beta_{\rm i}$ are constants,
  and $G(\hat{\eta})$ and $H(\hat{\eta})$ are functions of $\hat{\eta}$,
  all determined by the initial conditions.
  These are the super-horizon solutions.
\end{enumerate}
Combined with the initial conditions (\ref{GNNini}) and (\ref{GNsini}),
equations (\ref{GcNinf}) and (\ref{Gc0})  can be solved to yield the following results.
For clarity,
we shall denote $\hat{y}=1+A\hat{\eta}/2$ in $\widetilde{\cal G}^{N{\rm s}}$ and
$y_{\rm i}=1+A\eta_{\rm i}/2$ in $\widetilde{\cal G}^{NN'}_i$ both as $w$ :
\begin{eqnarray}
  \widetilde{\cal G}^{\rm cs}_{\infty} &=&
  \frac{1}{4A}(w^2\!-\!1)\left\{
    (3w^2\!-\!1)(3y^2\!-\!1)\log\left[\frac{(w+1)(y-1)}{(w-1)(y+1)}\right]
    \!-\!6(y-w)(3wy+1)
  \right\},
  \label{Gcsinf}\\
  \widetilde{\cal G}^{\rm cs}_0 &=&
  \frac {2(y^6w - w^6y - 5y^4w + 5w^4y + 15y^2w - 15yw^2 + 5w - 5y)}
    {5A (y^2 - 1)^2 (w^2 - 1)}\,,
  \label{Gcs0}\\
  \widetilde{\cal G}^{\rm cc}_{1\infty} &=&
  \frac{1}{2}(3y^2-1)(3w^2-2)-\frac{9}{2}wy(w^2-1)
  \nonumber\\
  &&+\frac{3}{4}w(w^2-1)(3y^2-1)
  \log\left[\frac{(y+1)(w-1)}{(y-1)(w+1)}\right],
  \label{Gcc1inf}\\
  \widetilde{\cal G}^{\rm cc}_{10}&=&
  \frac {2yw^5 -20yw^3 +20y^2w^2 +20w^2 -30yw -15y^4 -5 +3y^6 +25y^2}
  {5(y^2 - 1)^2(w^2 - 1)}  \, ,
  \label{Gcc10}\\
  \widetilde{\cal G}^{\rm cc}_{2\infty}&=&
  \widetilde{\cal G}^{\rm cs}_{\infty}\,,
  \label{Gcc2inf}\\
  \widetilde{\cal G}^{\rm cc}_{20}&=&
  \widetilde{\cal G}^{\rm cs}_0-\frac{4}{3}\widetilde{\cal G}^{\rm cr}_{20}\,,
  \label{Gcc20}\\
  \widetilde{\cal G}^{\rm cr}_{1\infty}&=& 0\,,
  \label{Gcr1inf}\\
  \widetilde{\cal G}^{\rm cr}_{10}&=&
  \frac {3(y^6-5y^2w^4+4yw^5+10y^2w^2 - 5y^4 - 5w^4+10y^2  - 20yw+ 10w^2)}
  {5(y^2 - 1)^2(w^2 - 1)^2}\,,
  \label{Gcr10}\\
  \widetilde{\cal G}^{\rm cr}_{2\infty}&=& 0\,,
  \label{Gcr2inf}\\
  \widetilde{\cal G}^{\rm cr}_{20} &=&
  \frac {-3}{10 A} \left[ 
    \frac {(y^2 + 4y + 5)(y - 1)^2}{(y + 1)^2}
    \log\left(\frac {y - 1}{w - 1}\right)
    + \frac {(y^2 - 4y + 5)(y + 1)^2}{(y - 1)^2}
    \log\left(\frac {w + 1}{y + 1}\right)
  \right.
  \nonumber \\
  +  &\!\!2 & \!\!
  \left.
    (w - y)
    \frac {(4yw^3 -6y^2w^2 -10w^2 +y^5w -4y^2w +7yw + 6y^2 - 4y^3 + 5+ y^4)}
    {(w^2 - 1)(y^2 - 1)^2}
  \right].
  \label{Gcr20}
\end{eqnarray}
We note that equations (\ref{Gcc2inf}),  (\ref{Gcc20}) and (\ref{Gcr2inf}) result directly
from the initial conditions (\ref{GNNini}) and (\ref{GNsini}).
They are consistent with equation (\ref{Gcs-2}).

\begin{figure}
  \centering\epsfig{figure=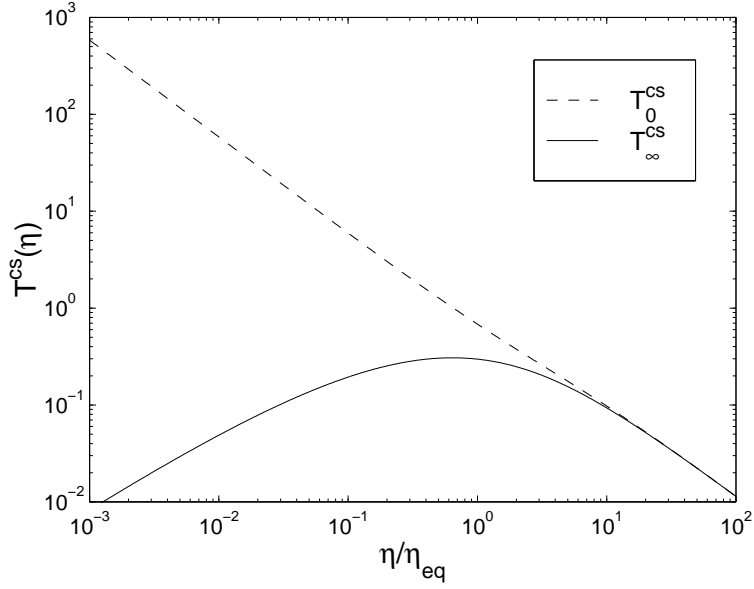, width=10cm}\\
  \caption[]
  {The source transfer functions
    $\widetilde{T}^{\rm cs}_0$ ($k\eta\ll 1$, dashed line) and
    $\widetilde{T}^{\rm cs}_{\infty}$ ($k\eta\gg 1$, solid line).
    }
  \label{fig-Tcs}
\end{figure}

\begin{figure}
  \centering\epsfig{figure=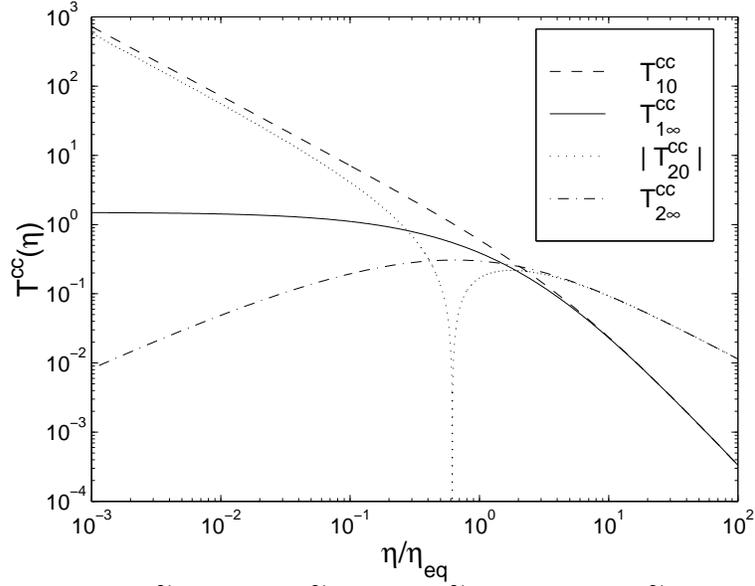, width=10cm}\\
  \caption[]
  {The source transfer functions
  $\widetilde{T}^{\rm cc}_{10}$ (dashed),
  $\widetilde{T}^{\rm cc}_{1\infty}$ (solid),
  $\widetilde{T}^{\rm cc}_{20}$ (dotted), and
  $\widetilde{T}^{\rm cc}_{2\infty}$ (dot-dashed).
  We have taken the absolute value of $\widetilde{T}^{\rm cc}_{20}$,
  because it becomes negative when $\eta\lesssim 0.6\eta_{\rm eq}$.
  }
  \label{fig-Tcc}
\end{figure}

\begin{figure}
  \centering\epsfig{figure=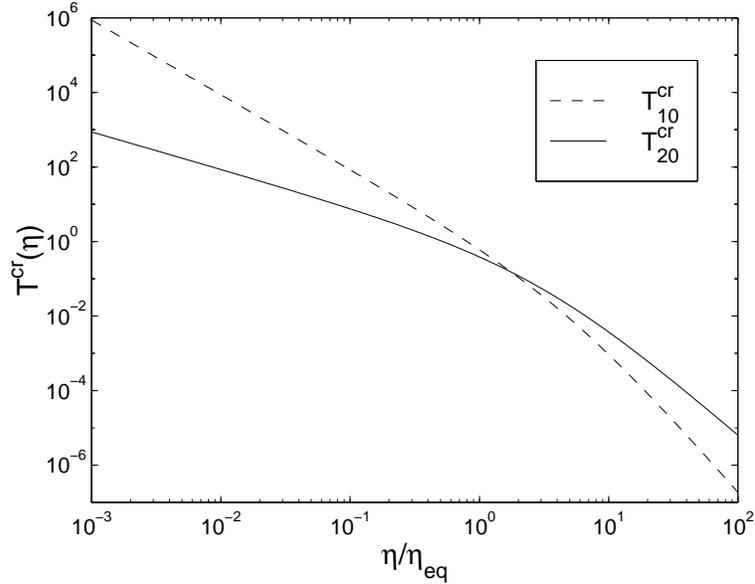, width=10cm}\\
  \caption[The source transfer functions
  $\widetilde{T}^{\rm cr}_{10}$ and
  $\widetilde{T}^{\rm cr}_{20}$.
  ]
  {The source transfer functions
    $\widetilde{T}^{\rm cr}_{10}$ (dashed) and
    $\widetilde{T}^{\rm cr}_{20}$ (solid).
    We note that
    $\widetilde{T}^{\rm cr}_{1\infty}
    =\widetilde{T}^{\rm cr}_{2\infty}=0$.
    }
  \label{fig-Tcr}
\end{figure}

Despite the complicated forms presented here,
all these Green functions have simple
asymptotic behaviors in the radiation- or matter-dominated regimes.
Since we are more interested in the matter perturbations today and
we know from equation (\ref{G_one}) that
$\widetilde{\cal G}^{\rm c} \propto \eta^2 \propto a$
when $\eta/\eta_{\rm eq} \rightarrow \infty$,
we can design a `source transfer function' as
\begin{equation}
  \label{T_limit}
  \widetilde{T}^{\rm c}(k; \hat{\eta})
  \equiv \lim_{\eta/\eta_{\rm eq} \rightarrow \infty}
  \frac{a_{\rm eq}}{a} \, \widetilde{\cal G}^{\rm c}(k; \eta, \hat{\eta})\, .
\end{equation}
Note that
this is different from the definition of the standard CDM transfer function
(\ref{cdm-transfer-function}).
Equations (\ref{Gcsinf})--(\ref{Gcr20}) then lead to
the source transfer functions:
\begin{eqnarray}
  \widetilde{T}^{\rm cs}_{\infty} &=&
  \frac{3}{4A}(w^2-1)\left[
    (3w^2-1)\log\left(\frac{w+1}{w-1}\right)
    -6w
  \right],
  \label{Tcsinf}\\
  \widetilde{T}^{\rm cs}_0 &=&
  \frac {2w}{5A  (w^2 - 1)}\,,
  \label{Tcs0}\\
  \widetilde{T}^{\rm cc}_{1\infty} &=&
  \frac{3}{2}(3w^2-2)
  + \frac{9}{4}w(w^2-1)
  \log\left(\frac{w-1}{w+1}\right),
  \label{Tcc1inf}\\
  \widetilde{T}^{\rm cc}_{10}&=&
  \frac {3}{5(w^2 - 1)}  \, ,
  \label{Tcc10}\\
  \widetilde{T}^{\rm cc}_{2\infty}&=&
  \widetilde{T}^{\rm cs}_{\infty}\,,
  \label{Tcc2inf}\\
  \widetilde{T}^{\rm cc}_{20}&=&
  \widetilde{T}^{\rm cs}_0-\frac{4}{3}\widetilde{T}^{\rm cr}_{20}
  =
  -\frac {2}{5 A} \left[ 
     \frac {w}{w^2 - 1}
    -\log\left(\frac {w + 1}{w - 1}\right)
  \right]
  \,,
  \label{Tcc20}\\
  \widetilde{T}^{\rm cr}_{1\infty}&=& 0\,,
  \label{Tcr1inf}\\
  \widetilde{T}^{\rm cr}_{10}&=&
  \frac {3}{5(w^2 - 1)^2}\,,
  \label{Tcr10}\\
  \widetilde{T}^{\rm cr}_{2\infty}&=& 0\,,
  \label{Tcr2inf}\\
  \widetilde{T}^{\rm cr}_{20} &=&
   \frac {3}{10 A} \left[ 
     \frac {2 w}{w^2 - 1}
    -\log\left(\frac {w + 1}{w - 1}\right)
  \right].
  \label{Tcr20}
\end{eqnarray}
We plot these source transfer functions in
Figures~\ref{fig-Tcs}, 
\ref{fig-Tcc},
and \ref{fig-Tcr}.
They are now only functions of the initial time, but not of the final time.
In the context of topological defects,
the defect source was formed at $\eta_{\rm i}\ll \eta_{\rm eq}$.
Therefore it would be also interesting to investigate the asymptotic behaviors
of the source transfer functions
with very early initial times.
For $\eta_{\rm i}\ll \eta_{\rm eq}$,
equations (\ref{Tcsinf})--(\ref{Tcr20}) become:
\begin{eqnarray}
  \widetilde{T}^{\rm cc}_{1\infty{(\rm i)}} = \frac{3}{2} \,,
  \quad
  \widetilde{T}^{\rm cs}_{\infty{(\rm i)}} =
  \widetilde{T}^{\rm cc}_{2\infty{(\rm i)}} =
  \frac{3\eta_{\rm i}}{2\eta_{\rm eq}}
  \log\frac{4\eta_{\rm eq}}{A\eta_{\rm i}} \, ,
  \quad
  \widetilde{T}^{\rm cr}_{1\infty{(\rm i)}}=
  \widetilde{T}^{\rm cr}_{2\infty{(\rm i)}}= 0\,,
  \label{Tcinf-i}\\
  \frac{3A}{2}\widetilde{T}^{\rm cs}_{0{(\rm i)}}=
  \widetilde{T}^{\rm cc}_{10{(\rm i)}}=
  -\frac{3A}{2}\widetilde{T}^{\rm cc}_{20{(\rm i)}}=
  A\frac{\eta_{\rm i}}{\eta_{\rm eq}} \widetilde{T}^{\rm cr}_{10{(\rm i)}}=
  A\widetilde{T}^{\rm cr}_{20{(\rm i)}} =
  \frac {3\eta_{\rm eq}}{5A\eta_{\rm i}} \, ,
  \label{Tc0-i}
\end{eqnarray}
where the subscript (i) denotes the condition $\eta_{\rm i}\ll \eta_{\rm eq}$.
These asymptotic behaviors can be clearly seen in
Figures~\ref{fig-Tcs}, \ref{fig-Tcc} and \ref{fig-Tcr}.
We note that on sub-horizon scales,
$\widetilde{T}^{\rm cs}_\infty$
has a maximum at $\eta\sim \eta_{\rm eq}$
as seen in Figure~\ref{fig-Tcs}.
Adding the fact that 
cosmic defects seed matter perturbations only on sub-horizon modes
due to the compensation mechanism,
it follows
that the defect-induced matter perturbations are seeded mainly
during the radiation-matter transition era.
This is a generically different mechanism from inflationary models,
in which matter perturbations are seeded during inflation
in the deep radiation era
when all the modes are well outside the horizon.
Nevertheless,
the defect and inflationary models both provide
scale-invariant perturbations at horizon crossing,
and 
these perturbations evolve similarly
after horizon crossing.

\subsection{Degeneracy of the Green functions}
\label{degeneracy-of-the-green-functions}

\begin{figure}
  \centering\epsfig{figure=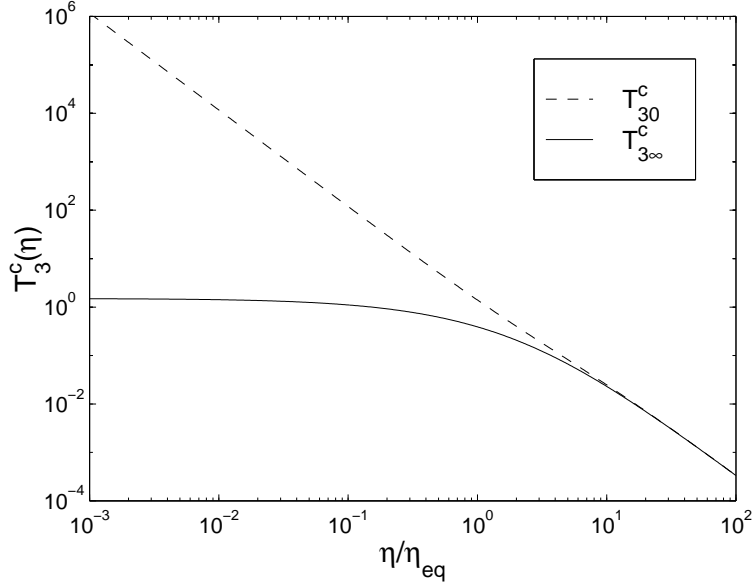, width=10cm}\\
  \caption[]
  {The source transfer functions
    $\widetilde{T}^{\rm c}_{30}$ (dashed line)
    and
    $\widetilde{T}^{\rm c}_{3\infty}$ ($=\widetilde{T}^{\rm cc}_{1\infty}$; solid line).
    }
  \label{fig-Tc3}
\end{figure}

In principle we need ten Green functions
(five for $\delta_{\rm c}=\delta_{\rm c}^{\rm I}+\delta_{\rm c}^{\rm S}$
and
five for $\delta_{\rm r}=\delta_{\rm r}^{\rm I}+\delta_{\rm r}^{\rm S}$)
in order to solve equations (\ref{delta-one}) and (\ref{delta-two})
by using the formalism (\ref{delta_I_N}) and (\ref{delta_S_N}).
However, in addition to the dependence (\ref{Gcs-2})
by which we can reduce the effective number of the Green functions by two,
there is another constraint we can invoke---the zero entropy fluctuation
on super-horizon scales in the initial conditions,
i.e.\ $s=\dot{s}=0$ at $\eta_{\rm i}$ for modes $k\ll 1/\eta_{\rm i}$
(see eqs.~[\ref{entropy}] and [\ref{initial-condition}]).
Since the formation time $\eta_{\rm i}$ of the active source is normally so early
that the condition $k\ll 1/\eta_{\rm i}$ (and thus $s=\dot{s}=0$)
is generally satisfied on the scales of our cosmological interest,
we can rewrite equation (\ref{delta_I_N}) as
\begin{equation}
  \label{delta_I_N-2}
  \widetilde{\delta}^{\rm I}_{N}({\bf k},\eta)
   = 
    \widetilde{\cal G}^{N}_3 (k;\eta,\eta_{\rm i})
    \widetilde{\delta}_{\rm c}({\bf k},\eta_{\rm i})
    +
    \widetilde{\cal G}^{N}_4 (k;\eta,\eta_{\rm i})
    \dot{\widetilde{\delta}}_{\rm c}({\bf k},\eta_{\rm i})
  \, ,  
\end{equation}
where
\begin{equation}
  \label{G_N_34}
  \widetilde{\cal G}^{N}_{i}
  =
  \widetilde{\cal G}^{N\rm c}_{i-2} +
  \frac{4}{3}\widetilde{\cal G}^{N\rm r}_{i-2} \,,
  \quad
  i=3, 4\,.
\end{equation}
From equations (\ref{Gcc1inf}), (\ref{Gcc10}),
(\ref{Gcr1inf}) and (\ref{Gcr10}),
we can get
\begin{eqnarray}
  \label{Gc30}
  \widetilde{\cal G}^{\rm c}_{3\infty}
  =\widetilde{\cal G}^{\rm cc}_{1\infty}, \quad
  \widetilde{\cal G}^{\rm c}_{30}
  & = &
  \left[5 (w^2 - 1)^2 (y^2 - 1)^{2}\right]^{-1}
  \left
    (y^{6} + 3y^{6}w^{2} - 5y^{4} - 15w^{2}y^{4}+ 15y^{2} 
  \right.
  \nonumber\\
  & &\left.
    + 45w^{2}y^{2} + 2w^{7}y - 10w^{3}y
    - 6yw^{5} - 50yw + 5 + 15w^{2}
  \right).
\end{eqnarray}
Using equation (\ref{Gcs-2}),
we can also obtain $\widetilde{\cal G}^{\rm c}_4=\widetilde{\cal G}^{\rm cs}$,
so that
\begin{equation}
  \widetilde{\cal G}^{\rm c}_{4\infty}=\widetilde{\cal G}^{\rm cs}_\infty, \quad
  \widetilde{\cal G}^{\rm c}_{40}=\widetilde{\cal G}^{\rm cs}_0.
\end{equation}
These results yield the source transfer functions:
\begin{equation}
  \label{Tc30}
  \widetilde{T}^{\rm c}_{3\infty}
  =\widetilde{T}^{\rm cc}_{1\infty}, \quad
  \widetilde{T}^{\rm c}_{30}=
  \frac{3w^2+1}{5(w^2-1)^2} \, , \quad
  \widetilde{T}^{\rm c}_{4\infty}
  =\widetilde{T}^{\rm cs}_{\infty}, \quad
  \widetilde{T}^{\rm c}_{40}
  =\widetilde{T}^{\rm cs}_{0}.
\end{equation}
If the initial time is deep in the radiation era,
i.e.\ $\eta_{\rm i}\ll\eta_{\rm eq}$, we further have
\begin{eqnarray}
  \widetilde{T}^{\rm c}_{30\rm(i)}
  = \frac{4 \eta_{\rm eq}^2}{5A^2\eta_{\rm i}^2}
  \propto \eta_{\rm i}^{-2},
  & \quad
  \widetilde{T}^{\rm c}_{3\infty\rm(i)}
  =\widetilde{T}^{\rm cc}_{1\infty\rm (i)}
  =\frac{3}{2} \propto  \eta_{\rm i}^0,
  \label{Tc3-propto}
  \\
  \widetilde{T}^{\rm c}_{40\rm(i)}
  =\widetilde{T}^{\rm cs}_{0\rm (i)}
  = \frac {2\eta_{\rm eq}}{5A^2\eta_{\rm i}}
  \propto \eta_{\rm i}^{-1},
  & \quad
  \widetilde{T}^{\rm c}_{4\infty\rm(i)}
  =\widetilde{T}^{\rm cs}_{\infty\rm (i)}
  =  \frac{3\eta_{\rm i}}{2\eta_{\rm eq}}
  \log\frac{4\eta_{\rm eq}}{A\eta_{\rm i}}
  \propto  \eta_{\rm i}\,,
  \label{Tc4-propto}
\end{eqnarray}
where the last proportionality is only an approximation.
Figure~\ref{fig-Tc3} shows the solutions of
$\widetilde{T}^{\rm c}_{30}$ and $\widetilde{T}^{\rm c}_{3\infty}$
($=\widetilde{T}^{\rm cc}_{1\infty}$),
while 
$\widetilde{T}^{\rm c}_{40}\equiv \widetilde{T}^{\rm cs}_{0}$
and 
$\widetilde{T}^{\rm c}_{4\infty}\equiv \widetilde{T}^{\rm cs}_{\infty}$
are already shown in Figure~\ref{fig-Tcs}.
We note the the asymptotic behaviors indicated in 
equations (\ref{Tc3-propto}) and (\ref{Tc4-propto})
can be clearly seen in Figures~\ref{fig-Tcs} and \ref{fig-Tc3}.
Therefore,
the original ten Green functions for solving
$\widetilde\delta_{\rm c}$ and $\widetilde\delta_{\rm r}$
have now been reduced to four functions:
two for $\widetilde\delta_{\rm c}$
($\widetilde{\cal G}^{\rm c}_4 \equiv \widetilde{\cal G}^{\rm cs}$
 and $\widetilde{\cal G}^{\rm c}_3$),
and two for $\widetilde\delta_{\rm r}$
($\widetilde{\cal G}^{\rm r}_4 \equiv \widetilde{\cal G}^{\rm rs}$
 and $\widetilde{\cal G}^{\rm r}_3$).
We shall concentrate only on the solutions of $\widetilde\delta_{\rm c}$,
while leaving those of $\widetilde\delta_{\rm r}$ elsewhere \cite{Wu2001}.
To calculate $\widetilde\delta_{\rm c}^{\rm S}$
we need $\widetilde{\cal G}^{\rm cs}$ using equation (\ref{delta_S_N});
to calculate $\widetilde\delta_{\rm c}^{\rm I}$
we need $\widetilde{\cal G}^{\rm c}_4=\widetilde{\cal G}^{\rm cs}$
and $\widetilde{\cal G}^{\rm c}_3$ using equation (\ref{delta_I_N-2}).
In solving $\widetilde\delta_{\rm c}^{\rm I}$,
we note that
$\widetilde{\cal G}^{\rm c}_3$ transfers the initial
perturbations of both matter and radiation
$[\widetilde\delta_{\rm c}(\eta_{\rm i})
+\widetilde\delta_{\rm r}(\eta_{\rm i})]$
to today,
while 
$\widetilde{\cal G}^{\rm cs}$ transfers the initial
perturbations of their time derivatives
$[\dot{\widetilde\delta}_{\rm c}(\eta_{\rm i})
+\dot{\widetilde\delta}_{\rm r}(\eta_{\rm i})]$
to the present.

\subsection{Solutions on intermediate scales}
\label{solutions-on-intermediate-scales}

\begin{figure}
  \centering\epsfig{figure=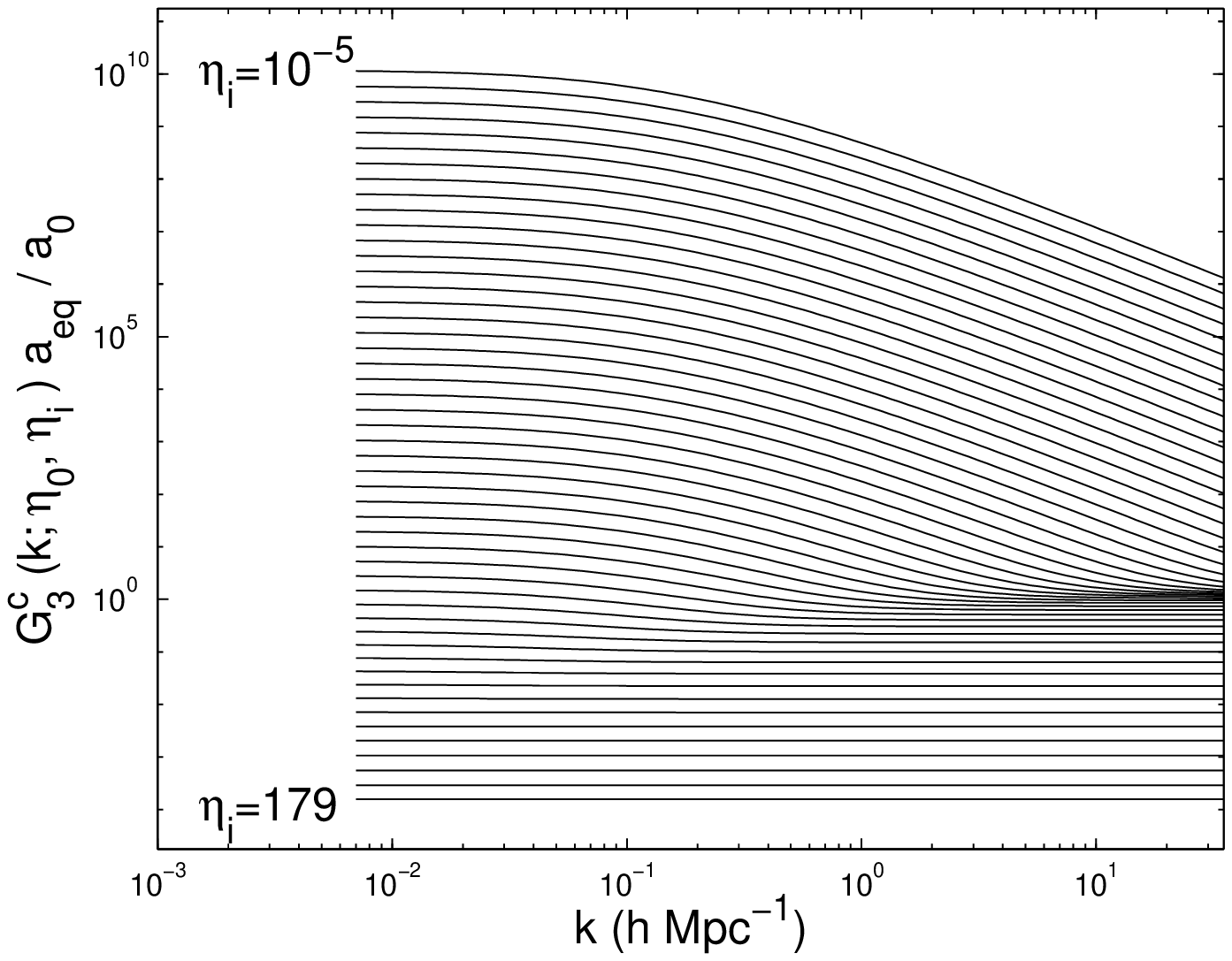, width=10cm}\\
  \centering\epsfig{figure=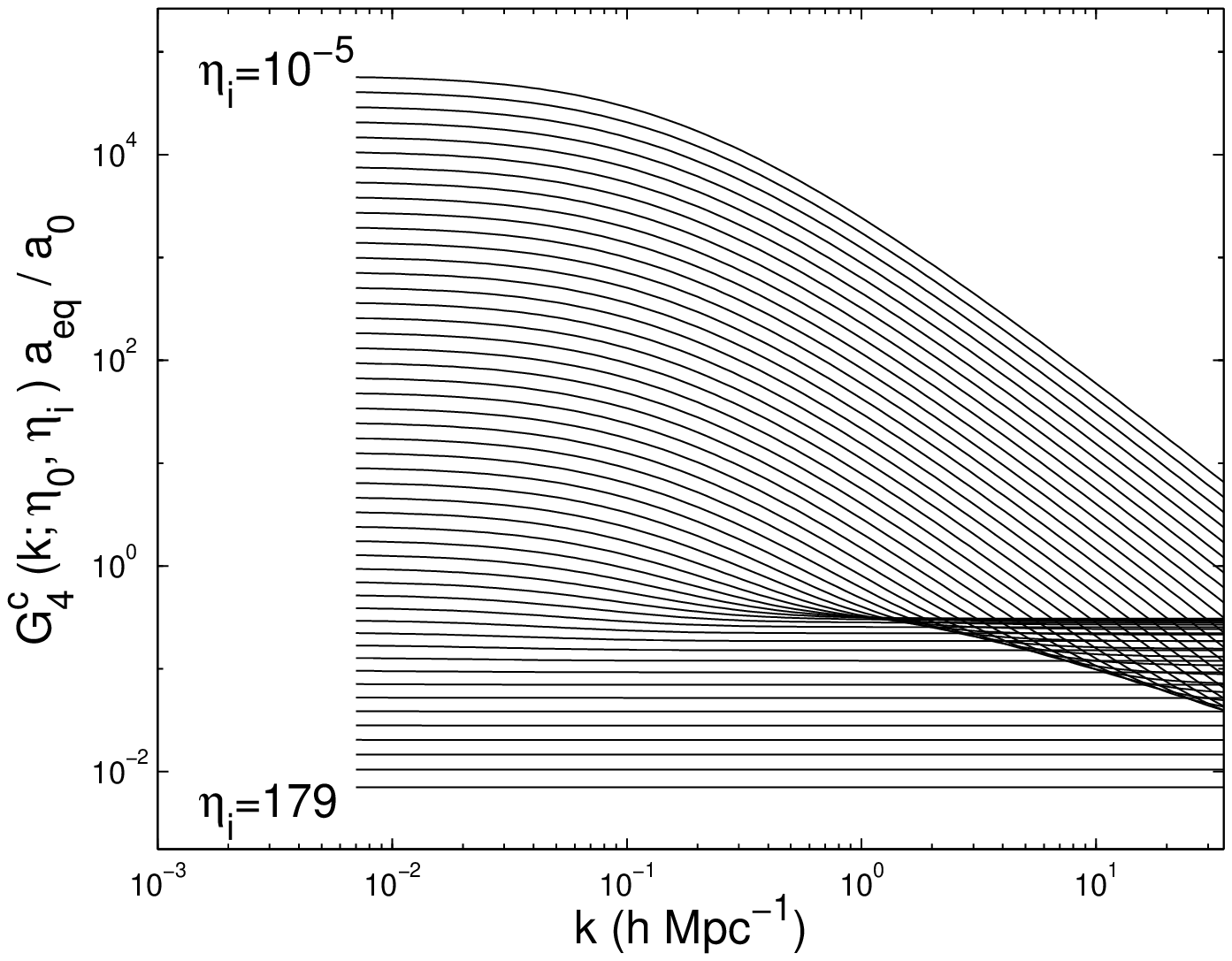, width=10cm}\\
  \caption[]
  {The numerical solutions of
    $\widetilde{\cal G}^{\rm c}_3(k; \eta_0,\eta_{\rm i})$ (upper panel)
    and $\widetilde{\cal G}^{\rm c}_4(k; \eta_0,\eta_{\rm i})$
    (=$\widetilde{\cal G}^{\rm cs}(k; \eta_0,\eta_{\rm i})$; lower panel).
    They both have been normalized to the scale factor today,
    $a_0/a_{\rm eq}$.
    Each line has a different initial time $\eta_{\rm i}$,
    whose smallest and largest values are labeled in both plots.
    Successive lines have even logarithmic time intervals,
    and $\eta_{\rm i}$ is in units of $\eta_{\rm eq}$.
    }
  \label{fig-Gc34}
\end{figure}

With $\widetilde{\cal G}^{\rm c}_3$ 
and $\widetilde{\cal G}^{\rm c}_4$ ($=\widetilde{\cal G}^{\rm cs}$)
as the two basis Green functions,
we can now work out the solutions on intermediate scales,
using results derived in previous sections.
In the matter era,
${\widetilde{\cal G}}^{\rm r}_i\ll {\widetilde{\cal G}}^{\rm c}_i$
on all scales
so from equation (\ref{G_one}) we know that 
${\widetilde{\cal G}}^{\rm c}_{i0}(k; \eta_0,\eta_{\rm i}) =
{\widetilde{\cal G}}^{\rm c}_{i\infty}(k; \eta_0,\eta_{\rm i})$
when $\eta_{\rm i} \gg \eta_{\rm eq}$.
This can be clearly seen from Figures~\ref{fig-Tcs}
and \ref{fig-Tc3}.
In the radiation era,
the perturbations
$[\widetilde\delta_{\rm c}(\eta_{\rm i})
+\widetilde\delta_{\rm r}(\eta_{\rm i})]$
or
$[\dot{\widetilde\delta}_{\rm c}(\eta_{\rm i})
+\dot{\widetilde\delta}_{\rm r}(\eta_{\rm i})]$
that were seeded well before the horizon crossing
will evolve in the same way as in the standard CDM model
due to the same zero entropy fluctuation initial condition.
Therefore the solution interpolating between
${\widetilde{\cal G}}^{\rm c}_{i0}(k; \eta_0,\eta_{\rm i})$
and
${\widetilde{\cal G}}^{\rm c}_{i\infty}(k; \eta_0,\eta_{\rm i})$
for $\eta_{\rm i}\ll \eta_{\rm eq}$
will be the standard CDM transfer function.
Thus we can write down a fit of the solution for
the full gamut of $k$ and $\eta_{\rm i}$ as
\begin{equation}
  \label{Gc}
  {\widetilde{\cal G}}^{\rm c}_i(k; \eta_0,\eta_{\rm i})
  =
  {\widetilde{\cal G}}^{\rm c}_{i\infty}(\eta_0,\eta_{\rm i})
  +
  \left[
  {\widetilde{\cal G}}^{\rm c}_{i0}(\eta_0,\eta_{\rm i})
  -{\widetilde{\cal G}}^{\rm c}_{i\infty}(\eta_0,\eta_{\rm i})
  \right]
  T(k)
  I(k;\eta_{\rm i}),
  \quad
  i=3,4 \,,
\end{equation}
where 
\begin{eqnarray}
  T(k)
  &=&
  \left[
    1 +
    \frac{(0.0534+\frac{2.75}{1+3.83k})k^2}{\ln(2e+0.11k)}
  \right]^{-1},
  \label{Tk-2}\\
  I(k; \eta_{\rm i})
  &=&
  \frac{1+30\eta_{\rm i}}
  {1+30\eta_{\rm i}(1+\frac{k\eta_{\rm i}}{2\pi})},
  \label{Iketa}
\end{eqnarray}
and
$k$ is in units of $\eta_{\rm eq}^{-1}$ (see equation (\ref{eta_eq})).
Here 
$T(k)\equiv T_{\rm c}(k,\eta_0; \Omega_{\rm B0}=0)$
is the standard CDM transfer function without baryons
(modified from Ref.~\cite{EisensteinHu};
see eq.~[\ref{cdm-transfer-function}] 
for the definition of $T_{\rm c}(k,\eta_0)$), 
and
$I(k; \eta_{\rm i})$ is a small correction 
near the horizon crossing
to make the analytic solutions (\ref{Gc}) fit the numerical results.
For a given mode which is initially outside the horizon,
the background contents of the universe
compensate the defect source until horizon crossing.
Therefore
the detailed behavior of these Green functions near the horizon scale
will affect the so-called compensation scale,
beyond which no perturbations can grow.
This means that
the correction function $I(k; \eta_{\rm i})$ in equation (\ref{Gc}) actually
plays an important role in getting the compensation scale right,
and we shall discuss this further 
in section~\ref{compensation-and-total-matter-perturbations}.
We have verified numerically for both
$\widetilde{\cal G}^{\rm c}_3$ and $\widetilde{\cal G}^{\rm c}_4$
that the fit (\ref{Gc}) is accurate within a $4\%$ error
for any $k$ and $\eta_{\rm i}$
(note that the initial conditions of
$\widetilde{\cal G}^{\rm c}_3$ and $\widetilde{\cal G}^{\rm c}_4$
in the numerical verifications
can be obtained from eqs.~[\ref{GNNini}], [\ref{GNsini}] and [\ref{G_N_34}]).
Figure~\ref{fig-Gc34} shows
the numerical solutions of 
$\widetilde{\cal G}^{\rm c}_3$ and $\widetilde{\cal G}^{\rm c}_4$
(=$\widetilde{\cal G}^{\rm cs}$)
within a chosen domain of $(k,\eta_{\rm i})$.
It confirms the asymptotic behaviors indicated by equations
in (\ref{Tc30}) 
(see also eqs.~[\ref{Tcsinf}], [\ref{Tcs0}] and [\ref{Tcc1inf}]),
and plotted in Figures~\ref{fig-Tcs} and \ref{fig-Tc3}.
The asymptotic behaviors shown by equations
(\ref{Tc3-propto}) and (\ref{Tc4-propto})
can be also marginally observed from Figure~\ref{fig-Gc34}.

\begin{figure}
  \centering\epsfig{figure=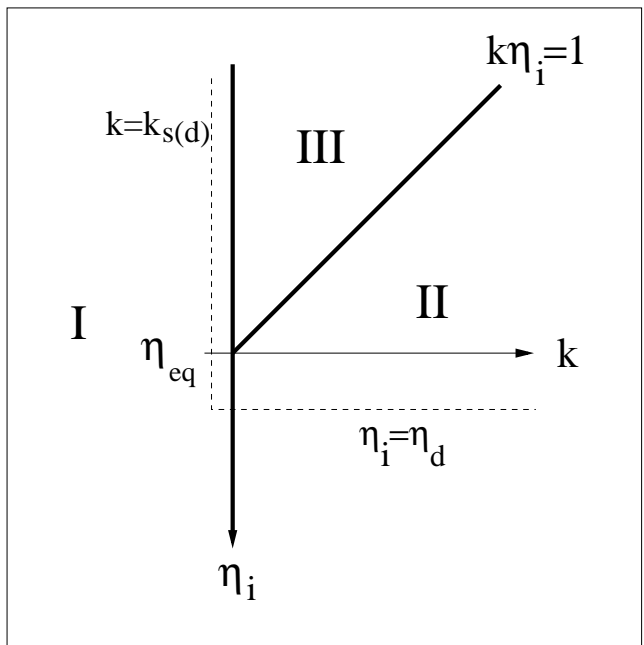, width=8.5cm}\\
  \caption[Three domains on the $(k, \eta_{\rm i})$-plane
  for the solutions of Green functions.
  ]
  {Three domains on the $(k, \eta_{\rm i})$-plane
    for the solutions of Green functions
    $\widetilde{\cal G}^{\rm c}_3$ and $\widetilde{\cal G}^{\rm c}_4$:
    Region I ($k<k_{\rm eq}=1/\eta_{\rm eq}$),
    Region II ($k>k_{\rm eq}$ and $k>1/\eta_{\rm i}$),
    and Region III ($k_{\rm eq}<k<1/\eta_{\rm i}$).
    These three regions are divided by the thick solid lines.
    Also shown are the $\eta_{\rm i}=\eta_{\rm d}$ (horizontal dashed line),
    and the $k=k_{\rm s(d)}$ (vertical dashed line).
    }
  \label{fig-k-eta}
\end{figure}

Schematically,
we can divide the $(k, \eta_{\rm i})$-plane into three regions
for the solutions of
$\widetilde{\cal G}^{\rm c}_i$ ($i=3,4$).
As shown in Figure~\ref{fig-k-eta},
these three domains are:
Region I ($k<k_{\rm eq}=1/\eta_{\rm eq}$),
Region II ($k>k_{\rm eq}$ and $k>1/\eta_{\rm i}$),
and Region III ($k_{\rm eq}<k<1/\eta_{\rm i}$).
In Region I, the solution of $\widetilde{\cal G}^{\rm c}_i$ is
$\widetilde{\cal G}^{\rm c}_{i0}$ because the horizon crossing happens
after $\eta_{\rm eq}$, after which
$\widetilde{\cal G}^{\rm c}_{i0}=\widetilde{\cal G}^{\rm c}_{i\infty}$
as argued before.
In Region II, the solution  is
$\widetilde{\cal G}^{\rm c}_{i\infty}$ because
all modes in this region are inside the horizon all the time.
We notice that $\widetilde{\cal G}^{\rm c}_{i0}$ merges with
$\widetilde{\cal G}^{\rm c}_{i\infty}$ at the boundary of Regions I and II,
where $\eta_{\rm i}>\eta_{\rm eq}$.
In Region III, the solution along the $k$ direction is in the same form as
the standard CDM transfer function.
This is because
modes with larger $k$ cross the horizon earlier,
so that their perturbations are suppressed after the horizon crossing 
for longer until $\eta_{\rm eq}$.
In addition,
the solution along the $\eta_{\rm i}$ direction
in Region III
is in the same form as $\widetilde{\cal G}^{\rm c}_{i0}$.
This is because 
modes in this region
are initially on super-horizon scales,
and 
a given mode with different initial time $\eta_{\rm i}$
will experience the same amount of suppression 
resulting from the period
between the horizon crossing and $\eta_{\rm eq}$.
Therefore, 
Regions I, II and III 
illustrate the intrinsic property of the solution (\ref{Gc}).

\subsection{The effect of baryons}
\label{the-effect-of-baryons}

There is one important issue which we have not discussed---the 
effect of baryons.
Prior to the photon-baryon decoupling at $\eta_{\rm d}$,
the CDM and baryons are dynamically independent.
In this era,
the photon-baryon fluid propagates as acoustic waves
with a sound speed given by equation~(\ref{c_gammaB}),
preventing baryons from collapsing on small scales.
Therefore there exists a sound horizon at the decoupling epoch $d_{\rm s(d)}$
(hereafter simply the sound horizon)
which is the distance such waves can travel prior to $\eta_{\rm d}$,
and
which is the largest scale at which the baryons can affect 
the evolution of density perturbations.
It has been shown that inside the sound horizon $d_{\rm s(d)}$,
not only 
are
the CDM perturbations seeded before $\eta_{\rm d}$
suppressed due to the presence of baryons
(e.g.~\cite{BonEfs,EisensteinHu,HuSugiyama2}),
but also 
the baryons themselves have an exponentially decaying power
due to the Silk damping \cite{Silk} (see also eq.~[\ref{damping-effects}]),
with acoustic oscillations due to the velocity overshoot
\cite{SunZel,PreVis}.
After the decoupling,
baryons evolve in the same way as the CDM does,
so 
the matter perturbations today
can be obtained by
linearly combining the CDM and baryonic fluctuations
at $\eta_{\rm d}$
(see section~\ref{instantaneous-decoupling} and 
eq.~[\ref{delta-m-decoupling}]),
and then evolving them to today.

It follows that  
the baryonic effects tend to suppress the matter perturbations
seeded before the decoupling epoch
($\eta<\eta_{\rm d}$, see the horizontal dashed line in Figure~\ref{fig-k-eta})
and on scales inside the sound horizon
(i.e.\ for $k>k_{\rm s(d)}\sim 1/d_{\rm s(d)}$,
see the vertical dashed line in Figure~\ref{fig-k-eta}).
The perturbations seeded after $\eta_{\rm d}$ or
on scales $k<k_{\rm s(d)}$ will not be affected by the baryons.
With this argument,
we can impose a suppression factor on our current solution (\ref{Gc}) 
to account for the effect of baryons,
i.e.\ the solution with the inclusion of baryons can be written as
\begin{equation}
  \label{GcB}
  {\widetilde{\cal G}}^{\rm c(B)}_i
  (k; \eta,\eta_{\rm i};h,\Omega_{\rm m0}, \Omega_{\rm B0})
  =
  {\widetilde{\cal G}}^{\rm c}_i(k; \eta,\eta_{\rm i}) 
  B(k, \eta_{\rm i};h,\Omega_{\rm m0}, \Omega_{\rm B0}),
\end{equation}
where $B(k, \eta_{\rm i};h,\Omega_{\rm m0}, \Omega_{\rm B0})$ 
accounts for the baryonic suppression:
\begin{equation}
  \label{B_k}
  B(k, \eta_{\rm i};h,\Omega_{\rm m0}, \Omega_{\rm B0})
  =
  \left\{
    \begin{array}{ll}
    \frac{T(k;h,\Omega_{\rm m0}, \Omega_{\rm B0})}
    {T(k;h, 1,0)}, 
    & 
    {\rm for}\ \eta_{\rm i} \ll \eta_{\rm d}\, , \\
    1,
    & 
    {\rm for}\ \eta_{\rm i} \gtrsim \eta_{\rm d}
    \textrm{ or } k<k_{\rm s(d)}\sim 1/d_{\rm s(d)},
    \end{array}
  \right.
\end{equation}
where $T(k;h,\Omega_{\rm m0}, \Omega_{\rm B0})$
is the usual standard CDM transfer function 
with the baryonic dependence.
One accurate fit of $T(k;h,\Omega_{\rm m0}, \Omega_{\rm B0})$
is provided in Ref.~\cite{EisensteinHu}.
We note that the ratio 
${T(k;h,\Omega_{\rm m0}, \Omega_{\rm B0})}/{T(k;h, 1, 0)}$
is unity outside the sound horizon 
($k<k_{\rm s(d)}\equiv 1/d_{\rm s(d)}$),
and is less than unity inside the sound horizon.
Referring to Figure~\ref{fig-k-eta}, 
equation (\ref{B_k}) means that
the value of $B(k, \eta_{\rm i};h,\Omega_{\rm m0}, \Omega_{\rm B0})$ 
is less than unity in the region to the right and above the dashed lines,
and is unity otherwise.
We also note that
in the low-$\Omega_{\rm m0}$ models, 
the sound horizon can be smaller than 
the radiation-matter equality horizon,
i.e.,
it is possible that
$k_{\rm s(d)}\equiv 1/d_{\rm s(d)}>k_{\rm eq}$ \cite{EisensteinHu}.
In addition,
there is a transition era ($\eta_{\rm i}\lesssim \eta_{\rm d}$)
which is not included in equation (\ref{B_k}).
This is because in this era 
the baryonic effects do not fully operate as in the regime 
$\eta_{\rm i} \ll \eta_{\rm d}$
so that a good fit is not trivial to obtain.
We have numerically verified equation (\ref{B_k}),
although
an accurate fit to the missing era $\eta_{\rm i}\lesssim \eta_{\rm d}$
has yet to be found.

\subsection{Solutions in $K\neq 0$ or $\Lambda\neq 0$ models}
\label{k-neq-0-or-l-neq-0-models}

The solutions we have obtained so far have assumed $K=\Lambda= 0$.
For $K\neq 0$ or $\Lambda\neq 0$,
the growing behavior of the CDM perturbations
departs from that of a flat $\Lambda=0$ model 
only at very late times in the matter era
(see later for a more detailed argument).
This allows us to apply a universal suppression factor on
${\widetilde{\cal G}}^{\rm c(B)}$ to account for 
the effects of curvature or $\Lambda$:
\begin{equation}
  \label{GcBol}
  {\widetilde{\cal G}}_i^{\rm c(B)}
  (k; \eta_0,\hat{\eta};h,\Omega_{\rm m0}, \Omega_{\rm B0},\Omega_{\Lambda 0})
  =
  \Omega_{\rm m0} {h}^2
  g(\Omega_{\rm m0},\Omega_{{\Lambda 0}})
  {\widetilde{\cal G}}_i^{\rm c(B)}
  (k; \eta_0,\hat{\eta}; 1,1, \Omega_{\rm B0},0),
\end{equation}
where $k$ is in units of $\Omega_{\rm m0}h^2 \, {\rm Mpc}^{-1}$,
and $g(\Omega_{\rm m0}, \Omega_{{\Lambda 0}})$ is given by \cite{Carroll}
\begin{equation}
  \label{gomega-0}
  g(\Omega_{\rm m0}, \Omega_{\Lambda 0})=\frac{5\Omega_{\rm m0}}
  {2\left[\Omega_{\rm m0}^{4/7}-\Omega_{\Lambda 0} +
      (1+\Omega_{\rm m0}/2)(1+\Omega_{\Lambda 0}/70)\right]}\ .
\end{equation}
In equation (\ref{GcBol}),
the leading factor $\Omega_{\rm m0} h^2$ results from the fact 
that the ratio of scale factors $a_0/a_{\rm eq}$ is
proportional to $\Omega_{\rm m0} h^2$ and
that the Green function
${\widetilde{\cal G}}_i^{\rm c(B)}={\widetilde{T}}_i^{\rm c(B)}a_0/a_{\rm eq}$
is proportional to this ratio.
The  factor $g(\Omega_{\rm m0}, \Omega_{\Lambda 0})$  accounts for
the suppression of the linear growth of density perturbations
in a $K\neq 0$ or $\Lambda$-universe
relative to an $\Omega_{\rm m0}=1$ and $ \Omega_{\Lambda 0}=0$
universe \cite{Carroll}
(also verified in Ref.~\cite{Einstein}).
The reason for $k$ to have the unit
$\Omega_{\rm m0}h^2 \, {\rm Mpc}^{-1}$ in equation (\ref{GcBol})
is that
the horizon size at radiation-matter equality $\eta_{\rm eq}$
is proportional to $(\Omega_{\rm m0} h^2)^{-1}$
(see eq.~[\ref{eta_eq}] in Appendix~\ref{background_cosmologies}).

For $K\neq 0$ or $\Lambda\neq 0$,
the extrapolation scheme (\ref{GcBol}) will be inaccurate 
when $\hat{\eta}$ is close to $\eta_0$,
i.e.,
when the background dynamics at $\hat{\eta}$ significantly
departs from that of a flat $\Lambda=0$ model.
Nevertheless,
this extrapolation scheme is still appropriate for most models with
active source for two reasons.
First,
in the context of cosmic defects,
the power of matter perturbations on the scales of our interest
($k \sim 0.01$--$1 h{\rm Mpc}^{-1}$) is mainly seeded around $\eta_{\rm eq}$
(see Figure~\ref{scalefactor} and the discussion 
after eq.~[\ref{Tc0-i}]).
At this time,
the curvature or $\Lambda$ effects are negligible.
Second,
at late times when the curvature or $\Lambda$  effects become important,
these scales of our interest are already well inside the horizon so that
any curvature terms in the perturbation equations can be neglected.
Therefore, the only required change in the perturbation equations
to account for the effects of curvature or $\Lambda$
is simply to incorporate the correct background dynamics,
and this involves only modifications in
$a(\eta)$, $\Omega_{\rm c}(\eta)$ and $\Omega_{\rm r}(\eta)$,
whose solutions are given in Appendix~\ref{background_cosmologies}.
As can be seen in Figure~\ref{scalefactor},
the presence of curvature or a cosmological constant
affects the background dynamics only at late times.
More precisely,
we verify that 
for
$(\Omega_{\rm m0},\Omega_{\Lambda 0})=
(0.2,0),(0.2,0.8),(1,0)$ and $(2.0,0)$,
the largest observable scale for matter perturbations
$k\approx 0.01 h{\rm Mpc}^{-1}$
corresponds to the horizon sizes
at $\eta \approx 5, 5, 27, 54\eta_{\rm eq}$ respectively,
whereas in these models
the curvature or cosmological-constant domination occurs only
at a much later epoch when $\eta>\eta_0$.
At these moments ($\eta \approx 5, 5, 27, 54\eta_{\rm eq}$),
the scale factor in the $K\neq 0$ or $\Lambda\neq 0$  models departs
from that in the flat $\Lambda=0$ model only by less than one percent.
Indeed,
we have numerically verified that
the extrapolation scheme (\ref{GcBol}) is accurate
within a 5\% error
for $\eta_{\rm i} \leq 60\eta_{\rm eq}$ and $\Omega_\Lambda \leq 0.85$
in $\Lambda$-models,
for $\eta_{\rm i} \leq 20\eta_{\rm eq}$ and $\Omega_{\rm m0} \geq 0.2$
in open $\Lambda=0$ models,
and for $\eta_{\rm i} \leq 200\eta_{\rm eq}$ and $\Omega_{\rm m0} \leq 2$
in closed $\Lambda=0$ models.
These ranges of cosmological parameters have apparently
covered the values of our interest.

\section{Important properties}
\label{important-properties}

With the Green-function solutions we have found,
we can now analytically investigate some important aspects
about the growth of cosmological matter perturbations.

\subsection{The standard CDM model}
\label{the-standard-cdm-model}

First we investigate the relationship
between our Green functions and the standard CDM transfer function,
and thereby
to justify the use of the standard CDM transfer function
in the analytic solution (\ref{Gc}).
In the standard CDM model,
there are no subsequent perturbations,
so we have 
$\widetilde{\delta}_N
=\widetilde{\delta}_N^{\rm I}+\widetilde{\delta}_N^{\rm S}
=\widetilde{\delta}_N^{\rm I}$.
As discussed in equations (\ref{rad-superh}) and (\ref{mat-superh}),
we also know that
the CDM perturbations have a growing mode
$\widetilde{\delta}_{\rm c}(k; \eta)\propto \eta^2$
on super-horizon scales  ($k\eta\ll 1$)
for $\eta\gg\eta_{\rm eq}$ or $\eta\ll\eta_{\rm eq}$.
For the super-horizon modes in the radiation era
and all modes in the matter era,
this allows us to write
\begin{equation}
  \label{A_j}
  \widetilde{\delta}_{\rm c}(k; \eta)=A_j(k) \eta^2, 
  \quad
  j= {\rm R,M}\, ,
\end{equation}
where $A_j$ is the coefficient of the growing mode
in the radiation era ($j=$R: $\eta\ll\eta_{\rm eq}$ and $k\eta\ll 1$)
or 
in the matter era ($j=$M: $\eta\gg\eta_{\rm eq}$).
Thus using our Green-function solutions (\ref{delta_I_N-2})
and (\ref{Gc})
with the initial conditions $s=\dot{s}=0$ and
$\dot{\widetilde{\delta}}_N(k; \eta_{\rm i})
=2\widetilde{\delta}_N(k; \eta_{\rm i})/\eta_{\rm i}$
as required by the adiabatic inflationary model,
we can derive the standard CDM transfer function as
\begin{eqnarray}
  \frac{A_{\rm M}}{A_{\rm R}}
  & = &
  \frac{\widetilde{\delta}_{\rm c}^{\rm I}(k; \eta)\eta_{\rm i}^2}
  {\widetilde{\delta}_{\rm c}(k; \eta_{\rm i})\eta^2}
  =
  \frac{A^2\eta_{\rm i}^2a_{\rm eq}}{4\eta_{\rm eq}^2a}
  \left[
    \widetilde{\cal G}^{\rm c}_{3}+
    \frac{2}{\eta_{\rm i}}
    \widetilde{\cal G}^{\rm c}_{4}
  \right]
  \nonumber\\
  & = &
  \frac{1}{4\eta_{\rm eq}^2}
  \left[
    A^2\eta_{\rm i}^2\widetilde{T}^{\rm c}_{30\rm(i)}+
    2A^2\eta_{\rm i} \widetilde{T}^{\rm c}_{40\rm(i)}
  \right]T(k)
  =\frac{2}{5}T(k)\,,
  \label{transTeta}
\end{eqnarray}
where we have used $\eta_{\rm i}\ll\eta_{\rm eq}\ll\eta$
and equations 
(\ref{aflat}), (\ref{T_limit}), (\ref{Tc3-propto}) and (\ref{Tc4-propto}),
and the last expression was obtained based on the formalism (\ref{Gc}).
First,
we note that
the two terms involving $\widetilde{T}^{\rm c}_{30\rm(i)}$
and $\widetilde{T}^{\rm c}_{40\rm(i)}$ are equal,
meaning that
the two sets of initial perturbations
$[\widetilde\delta_{\rm c}(\eta_{\rm i})
+\widetilde\delta_{\rm r}(\eta_{\rm i})]$
and
$[\dot{\widetilde\delta}_{\rm c}(\eta_{\rm i})
+\dot{\widetilde\delta}_{\rm r}(\eta_{\rm i})]$
contribute equally to the present matter perturbations.
Second,
 the $T(k)$ here is nothing but 
the standard CDM transfer function which we have defined earlier.
Third,
the coefficient $2/5$ in the final result of equation (\ref{transTeta})
is well known (e.g.~\cite{turok,BonEfs}),
and here we obtained it using our Green-function solutions.
This coefficient can be also obtained by
first knowing from equation (\ref{density-eqn3}) that 
$\tau_{00}$ is a constant on super-horizon scales ($k\ll 1/\eta$),
and then 
using its definition (\ref{tau00}) and equation (\ref{A_j})
to compare its expressions for $j=$R, M.
One will find
$\tau_{00}=A_{\rm R}/\pi G=5A_{\rm M}/2\pi G$,
which implies $A_{\rm M}/A_{\rm R}=2/5$ for $k\ll 1/\eta$.
Thus the above derivation and result
not only illustrate the relation between our Green functions and
the the standard CDM transfer function $T(k)$,
but also justify the use of $T(k)$ in our formalism (\ref{Gc}).

\subsection{Independence of the initial conditions}
\label{independence-of-the-initial-conditions}

One important problem for structure formation with causal seeds
is to investigate
how the source energy was compensated into the radiation and matter background
when the seeds were formed at $\eta_{\rm i}$.
From the result (\ref{tau00-propto})
we know that the power spectrum of the pseudo energy $\widetilde{\tau}_{00}$
should decay as $k^4$ on super-horizon modes.
As argued in equation (\ref{initial-condition}),
we can thus take $\widetilde{\tau}_{00}=0$ as part of the initial conditions
provided that
the scales of interest are well outside the horizon initially.
For similar reasons we can take $\widetilde{s}=\dot{\widetilde{s}}=0$,
where $s=3\delta_{\rm r}/4-\delta_{\rm c}$.
In addition, from equation (\ref{tau00}) without baryons, we have
\begin{equation}
  \label{tau00-2}
  \tau_{00}=\Theta_{00}
  +\frac{3}{8\pi G}\left(\frac{\dot{a}}{a}\right)^2
  \sum_{N={\rm c,r}} \Omega_N\delta_N
  +\frac{1}{4\pi G}\frac{\dot{a}}{a}\dot{\delta_{\rm c}}.
\end{equation}
Since $\tau_{00}=0$ is required at $\eta_{\rm i}$, 
it follows that for a given $\Theta_{00}({\bf x}, \eta_{\rm i})$,
one can have different ways of compensating it into
between $\delta_N$ and $\dot{\delta}_N$.
It is thus vital to check the dependence of
the resulting $\delta^{\rm I}_{\rm c}(\eta)$
on the way we compensate $\Theta_{00}({\bf x}, \eta_{\rm i})$ into
the background initially.
Consider the following two extreme cases,
both satisfying $\widetilde{\tau}_{00}=\widetilde{s}=\dot{\widetilde{s}}=0$
on super-horizon scales at $\eta_{\rm i}$:
\begin{enumerate}
\item
  \label{case1}
  $\widetilde{\delta}_{\rm c}=3\widetilde{\delta}_{\rm r}/4=0$,
  $\dot{\widetilde{\delta}_{\rm c}}=3\dot{\widetilde{\delta}_{\rm r}}/4=
  [-4\pi G (a/\dot{a}) \widetilde{\Theta}_{00}]_{\rm i}$:
  Using equation (\ref{delta_I_N-2}),
  the normalized resulting initial perturbations can be calculated as
  \begin{equation}
    \label{delta1}
    \Delta_1(\eta,\eta_{\rm i})
    =
    \frac{\widetilde{\delta_{\rm c}}^{\rm I}(k, \eta)}
    {[-4\pi G (a/\dot{a})\widetilde{\Theta}_{00}]_{\rm i}}
    =
    \widetilde{\cal G}^{\rm c}_4
    =
    \widetilde{\cal G}^{\rm cs}\,.
  \end{equation}
\item
  \label{case2}
  $\widetilde{\delta}_{\rm c}=3\widetilde{\delta}_{\rm r}/4=
  [-8\pi G(a/\dot{a})^2\widetilde{\Theta}_{00}/(4-\Omega_{\rm c})]_{\rm i}$,
  $\dot{\widetilde{\delta}_{\rm c}}=3\dot{\widetilde{\delta}_{\rm r}}/4=0$:
  Similarly we have
  \begin{equation}
    \label{delta2}
    \Delta_2(\eta,\eta_{\rm i})
    =
    \frac{\widetilde{\delta_{\rm c}}^{\rm I}(k, \eta)}
    {[-4\pi G (a/\dot{a})\widetilde{\Theta}_{00}]_{\rm i}}
    =
    \widetilde{\cal G}^{\rm c}_{3}
    \frac{2w(w^2-1)}{A(3w^2+1)} \,.
  \end{equation}
\end{enumerate}
To see the difference in $\widetilde{\delta_{\rm c}}^{\rm I}(k, \eta_0)$ today
resulting from these two cases,
one can calculate
\begin{equation}
  \label{D12}
  D_{12}(\eta_0,\eta_{\rm i})
  =
  \frac{\Delta_2}{\Delta_1}-1
  =
  \frac{2w(w^2-1)}{A(3w^2+1)}
  \frac{\widetilde{T}^{\rm c}_{30}}{\widetilde{T}^{\rm cs}_{0}}
  =0 \,,
\end{equation}
where we have used equations  (\ref{Tcs0}), (\ref{Tc30}) and (\ref{Gc}).
This implies that no matter how the source $\Theta_{00}({\bf x}, \eta_{\rm i})$
is compensated into the background when it was formed
(i.e.\ with any portions between $\delta_N$ and $\dot{\delta}_N$ initially),
it results in the same $\widetilde{\delta_{\rm c}}^{\rm I}(k, \eta_0)$ today
on scales which were outside the horizon at $\eta_{\rm i}$.
We note that
this independence of the initial conditions was first numerically observed
in Ref.~\cite{turok},
and here we have provided an analytic proof.

\subsection{Compensation and total matter perturbations}
\label{compensation-and-total-matter-perturbations}

\begin{figure}
  \centering\epsfig{figure=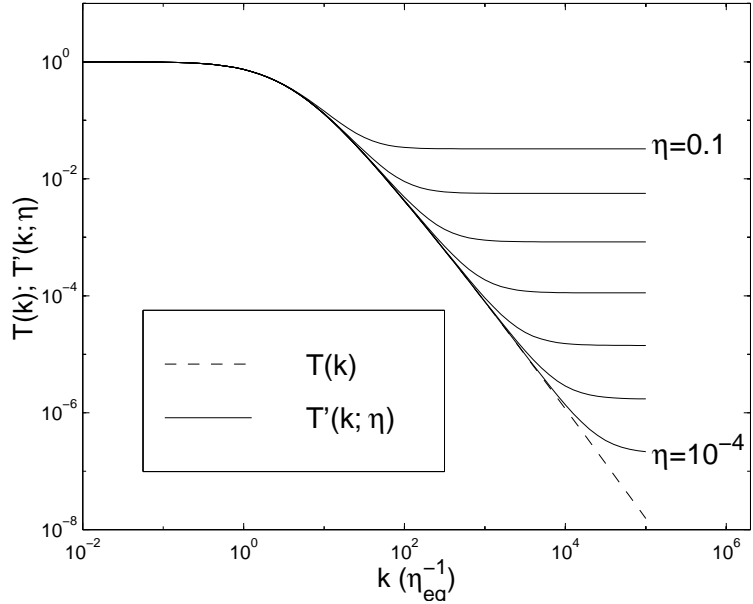, width=10cm}\\
  \caption[The function ${T'}(k; \eta)$ and
  and the standard CDM transfer function ${T}(k)$.]
  {The function ${T'}(k; \eta)$ (solid lines)
    and the standard CDM transfer function ${T}(k)$
    (the dashed line).
    Each solid line has different $\eta$,
    whose highest and lowest values are labeled in units of $\eta_{\rm eq}$.
    Successive lines have even logarithmic time intervals.
    }
  \label{fig-Tk-prime}
\end{figure}

With a complete set of Green functions for both initial and subsequent
perturbations, we can now investigate the resulting total CDM perturbations
and therefore the compensation mechanism in models with active source.
Having seen the independence of the resulting
$\widetilde{\delta_{\rm c}}^{\rm I}({\bf k}, \eta_0)$ on the way
the source energy is initially compensated into various background components,
we can invoke equation (\ref{delta1}) for
$\widetilde\delta^{\rm I}_{\rm c}$, and
equation (\ref{delta_S_N}) for
$\widetilde\delta^{\rm S}_{\rm c}$ to obtain
$  \widetilde{\delta}_{\rm c}({\bf k},\eta_0)=
\widetilde{\delta}_{\rm c}^{\rm I}({\bf k}, \eta_0)+
\widetilde{\delta}_{\rm c}^{\rm S}({\bf k}, \eta_0)$.
For a given mode at which $k\eta_{\rm i}\ll 1$ initially, we have:
\begin{eqnarray}
  \widetilde{\delta}_{\rm c}({\bf k},\eta_0)
  & = &
  \widetilde{\delta}_{\rm c}^{\rm I}({\bf k}, \eta_0)+
  \widetilde{\delta}_{\rm c}^{\rm S}({\bf k}, \eta_0)
  \nonumber \\
  & = &
  4\pi G
  \left[
    -
    \frac{a(\eta_{\rm i})}{\dot{a}(\eta_{\rm i})}
    \widetilde{\cal G}^{\rm cs}(k; \eta_0,\eta_{\rm i})
    \widetilde\Theta_{00}({\bf k}, \eta_{\rm i})    
    +
    \int_{\eta_{\rm i}}^{\eta_0}
    \widetilde{\cal G}^{\rm cs}(k; \eta_0,\hat{\eta})
    \widetilde\Theta_+({\bf k}, \hat{\eta})
    d\hat{\eta}
  \right]
  \label{delta_comp1}\\
  & = &
  \frac{8\pi G a_0}{5A^2 a_{\rm eq}}
  \left[
    -
    T(k)
    \widetilde\Theta_{00}({\bf k}, \eta_{\rm i})    
    +
    \int_{\eta_{\rm i}}^{\eta_0}
    {T'}(k; \hat{\eta})
    \frac{\dot{a}(\hat{\eta})}{a(\hat{\eta})}
    \widetilde\Theta_+({\bf k}, \hat{\eta})
    d\hat{\eta}
  \right]
  \label{delta_comp2}\\
  & = &
  \frac{8\pi G a_0}{5A^2 a_{\rm eq}}
  \left\{
    -
    T(k)
    {\widetilde\Theta_{00}({\bf k}, \eta_0)}
    +
  \right.
  \nonumber\\
  & &
  \left.
    \int_{\eta_{\rm i}}^{\eta_0}
    \left[
      {T'}(k; \hat{\eta})
      \frac{\dot{a}(\hat{\eta})}{a(\hat{\eta})}
      \widetilde\Theta_+({\bf k}, \hat{\eta})
      +
      T(k)
      \dot{\widetilde\Theta}_{00}({\bf k}, \hat{\eta})     
      \right]
    d\hat{\eta}
  \right\},
  \label{delta_comp3}
\end{eqnarray}
where
\begin{equation}
  \label{Tk-prime}
  {T'}(k; \hat{\eta})
  =
  \frac{\widetilde{\cal G}^{\rm c}_4(k; \eta_0, \hat\eta)}
  {\widetilde{\cal G}^{\rm c}_{40}(k; \eta_0, \hat\eta)}\,,
\end{equation}
and $\widetilde{\cal G}^{\rm c}_4(k; \eta_0, \hat\eta)$ is given by (\ref{Gc}).
The function ${T'}(k; \hat{\eta})$ is plotted in Figure~\ref{fig-Tk-prime}.
Here we notice that
the quantities inside the outer most brackets in equations 
(\ref{delta_comp2}) and (\ref{delta_comp3})
are equivalent to nothing but the coefficient 
of the growing mode in CDM perturbations.
Using equation (\ref{delta_comp2}),
one can obtain the resulting perturbations
$\widetilde{\delta}_{\rm c}({\bf k},\eta_0)$
by knowing the initial $\widetilde\Theta_{00}({\bf k}, \eta_{\rm i})$
and integrating the evolution history of 
$\widetilde\Theta_+({\bf k}, \hat{\eta})$.
Hence this expression is convenient for numerical purposes.
In addition, we see that
the first term in equation (\ref{delta_comp2})
comes simply from the initial source energy,
serving with an opposite sign to account for energy conservation.
This is the so-called compensation.
On the other hand, the second term results from the subsequent
evolution of $\widetilde\Theta_+({\bf k}, \hat{\eta})$,
which actively creates the CDM density perturbations
on sub-horizon scales (see later).
This term also provides a way for defects to create non-Gaussianity.

Alternatively,
equation (\ref{delta_comp3}) provides a both physically
and mathematically transparent way
of interpreting how the perturbations are seeded by the source.
First consider the integral term for a given mode $k$.
When the mode is well outside the horizon, i.e.\ $\hat\eta \ll 1/k$,
${T'}(k; \eta)$ equals $T(k)$ by definition.
Hence the two terms inside the inner brackets reduce to
$\widetilde\Theta_{0i,i}({\bf k}, \hat{\eta})T(k)$ due to
source stress-energy conservation (\ref{conserve-seed1}).
Since the power spectrum of $\widetilde\Theta_{0i,i}({\bf k}, \hat{\eta})$
falls off as $k^4$ outside the horizon (see eq.~[\ref{theta0i-propto}]),
we expect the quantity inside the brackets to be negligible until
the given mode approaches horizon crossing.
Near horizon crossing,
$\widetilde\Theta_{0i,i}({\bf k}, \hat{\eta})$ is no longer small,
and 
${T'}(k; \eta)$ starts departing from $T(k)$
(i.e.\ ${T'}(k; \eta)\approx$ constant $> T(k)\propto k^{-2}$,
see Figure~\ref{fig-Tk-prime}),
so the two terms inside the inner brackets begin to contribute to the integral.
This also explains
why the correction function $I(k; \eta_{\rm i})$ in equation (\ref{Gc})
is important in affecting the compensation scale.
After horizon crossing,
the significance of the two terms inside the inner brackets then depends on 
the subhorizon behaviours
of their power spectra.

As for the first term in equation (\ref{delta_comp3}),
we see that 
for a superhorizon mode today,
the integral in (\ref{delta_comp3}) is negligible as argued above
so that only the first term contributes.
It serves to give the opposite sign to the source energy
so as to account for energy conservation 
on superhorizon scales today,
and thus for the compensation
at the present epoch $\eta_0$.
On the other hand,
if a given  mode is well inside the horizon today,
then the first term will be negligible 
provided that 
the source energy $\widetilde\Theta_{00}({\bf k}, \eta_0)$
has a power-law fall-off inside the horizon,
as it does for cosmic strings.
Therefore in calculating CDM perturbations
on scales of our interest,
which are well inside the horizon today,
the first term in equation (\ref{delta_comp3}) is negligible,
so that 
it will not affect 
our compensation argument observed from the integral.

This argument can be further strengthened by deriving the pseudo-energy today.
From the definition of $\tau_{00}$ (\ref{tau00-2}) and
the final result of equation (\ref{delta_comp3}),
one obtains
\begin{eqnarray}
  \widetilde{\tau}_{00}({\bf k},\eta_0)
  &=& 
  \left(1-T(k)_{ }^{ }\right)
  {\widetilde\Theta_{00}({\bf k}, \eta_0)}
  +
  \nonumber \\
  & &
  \int_{\eta_{\rm i}}^{\eta_0}
  \left[
      {T'}(k; \hat{\eta})
      \frac{\dot{a}(\hat{\eta})}{a(\hat{\eta})}
      \widetilde\Theta_+({\bf k}, \hat{\eta})
      +
      T(k)
      \dot{\widetilde\Theta}_{00}({\bf k}, \hat{\eta})     
  \right]
  d\hat{\eta}\, .
  \label{tau00_0}
\end{eqnarray}
From this result,
one can clearly see that
for super-horizon modes,
$T(k)$ is unity by definition so that only the integral survives.
We have also seen from an earlier argument that
on super-horizon scales,
the quantity inside the square brackets is nothing but the 
$\widetilde\Theta_{0i,i}({\bf k}, \hat{\eta})$,
which has a $k^4$ fall-off power spectrum (see eq.~[\ref{theta0i-propto}]).
It follows immediately from equation (\ref{tau00_0}) that
the pseudo-energy today, $\tau_{00}(\eta_0)$, 
has a $k^4$-decay power spectrum outside the horizon.
This result confirms the super-horizon behavior of $\tau_{00}$
presented in equation (\ref{tau00-propto}).
On the other hand,
although $(1-T(k))$ 
is approximately unity for sub-horizon modes,
the usual sub-horizon power-law decay in 
$\widetilde\Theta_{00}({\bf k}, \eta_0)$ 
(as in the case of cosmic strings)
will still make the first term in equation (\ref{tau00_0}) negligible
inside the horizon.

Thus we can see explicitly in a  neat mathematical form
how compensation acts on a given length-scale.
From this analysis
we can also see that the compensation scale is determined 
not only by the functions ${T'}(k; \eta)$ and $T(k)$, 
but also by the properties of the source near
the horizon scale.
Once the detailed behavior of the source near the horizon scale is known,
we can accurately locate the compensation scale 
using equation (\ref{delta_comp3}) or (\ref{tau00_0}).
We note that
this result is different from the claim in Ref.~\cite{painless},
where
multi-fluid compensation back-reaction effects were studied to show that
the compensation scale arises naturally and uniquely from
an algebraic identity in the perturbation analysis.
Ref.~\cite{RobWan} also investigated the compensation scale,
and
found constraints on the generation of super-causal-horizon
energy perturbations from a smooth initial state,
under a simple physical scheme.
The compensation wavenumber was found to be constrained with
$k_{\rm c}\gtrsim 2 \eta^{-1}$ due to causality,
depending on the behavior of the causal events.
This result is not inconsistent with our finding above,
where we further provide a quantitative way
to locate the compensation scale for any given specific model.

\section{Summary and Conclusion}
\label{matpert-conclusion} 

In this paper
we present a formalism which can be used to study the evolution
of cosmological perturbations in the presence of causal seeds.
In this formalism we invoked the fluid approximation in the synchronous gauge 
to model the contents of the universe, 
and assumed photon-baryon tight coupling until the last-scattering epoch
to account for the baryonic effects.
The approximation of instantaneous decoupling of photons and baryons
 was then employed at  the last-scattering epoch.
In particular,
we demonstrated the accuracy of our formalism
in the context of the standard CDM model,
by comparing our results
of density perturbations 
with those calculated from CMBFAST.

We then derived 
the analytic solutions
of matter density perturbations 
in a flat $\Lambda=0$ cosmology.
The errors in Ref.~\cite{VeeSte} were corrected
to yield a complete set of Green-function solutions
for the super-horizon and sub-horizon modes
(eqs.~[\ref{deltaN}], [\ref{delta_I_N}], [\ref{delta_S_N}],
[\ref{Gcsinf}]--[\ref{Tc0-i}]).
The degeneracy among these Green functions
was then found by comparing their initial conditions 
and employing the zero-entropy initial condition
(eqs.~[\ref{Gcs-2}], [\ref{delta_I_N-2}]).
This effectively reduces the number of the Green functions needed
in the perturbation solutions
(eqs.~[\ref{deltaN}], [\ref{delta_S_N}], 
[\ref{delta_I_N-2}]--[\ref{Tc4-propto}]).
With this great simplification,
the solutions on intermediate scales
were then easily found by the use of the standard CDM transfer function
(eq.~[\ref{Gc}]).
This complete set of solutions were numerically verified to high accuracy.
The baryonic effects were also considered
(eq.~[\ref{GcB}]).
We then extrapolated
these Green-function solutions to $K\neq 0$ or $\Lambda\neq 0$ models
(eq.~[\ref{GcBol}]),
with numerical justifications to high accuracy.

Using these Green-function solutions,
we investigated several important aspects of structure formation
with causal source.
We first demonstrated the relation between our Green functions
and the standard CDM transfer function
(eq.~[\ref{transTeta}]).
Second
we proved that the resulting matter perturbations today is independent of
the way the source was initially compensated into the background contents
of the universe
(eq.~[\ref{D12}]).
With our Green-function solutions and 
the use of the pseudo-stress-energy tensor,
we finally addressed the compensation mechanism
in a mathematically and physically explicit way
(eqs.~[\ref{delta_comp2}], [\ref{delta_comp3}], [\ref{tau00_0}]).
In particular,
the compensation scale was shown to be dependent not only on the dynamics
of the universe,
but also on the properties of the source near the horizon scale.
Once given the detailed behavior of the source near the horizon scale,
the compensation scale can be accurately located using our Green functions
(eq.~[\ref{tau00_0}]).

Although in the literature,
there have been detailed treatments of theories with causal seeds,
the formalism and its analytic solutions presented here 
will provide not only a physically transparent way for understanding
the evolution of matter perturbations,
but also a computationally economical scheme
which is particularly pertinent
when one needs to investigate the phase information
of the resulting cosmological perturbations. 
Following the same line of development,
we have been also working on the analytic solutions
for radiation perturbations \cite{Wu2001},
which will be useful in computing the full-sky CMB anisotropies
seeded by topological defects.
Finally, 
we note that
although we have been concentrating on investigating the perturbations
with causal source,
our Green-function solutions are completely general and therefore can
be also applied to the study of models with acausal source.


{\bf {Acknowledgments}}---
We thank
Andrew Liddle,
Paul Shellard, Radek Stompor,
and
Neil Turok
for useful conversations.
We acknowledge support from 
NSF KDI Grant (9872979) and
NASA LTSA Grant (NAG5-6552).



\appendix
\section{Cosmological background dynamics}
\label{background_cosmologies}

With the discovery of the CMBR in 1964 \cite{Penzias},
the universe is believed to be
mainly composed of not only matter but also radiation.
After the discovery,
several authors worked out the solutions in some FRW models
with both radiation and matter \cite{Jac,Coh,Che,McI}.
In this appendix,
we aim to derive the general solution of FRW models,
in the presence of both curvature and a cosmological constant.

\begin{figure}
  \centering\epsfig{figure=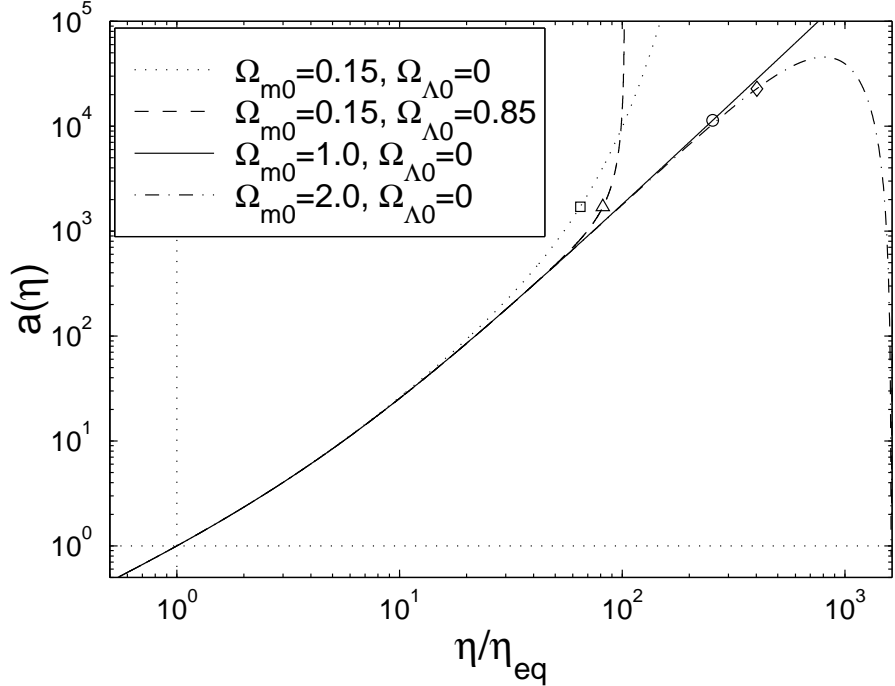, width=120mm}
  \caption[]
  {The evolution of background dynamics in various cosmologies.
    Plotted are exact solutions of the scale factor $a(\eta)$. 
    The square, triangle, circle and diamond mark the universe today
    for different models,
    each with $H_0=70$ km s$^{-1}$Mpc$^{-1}$.
    }
  \label{scalefactor}
\end{figure}

We assume that the universe is homogeneous and isotropic,
and is filled with two fluids, radiation and dark matter, 
whose stress-energy tensors are also homogeneous and isotropic on average.
We shall ignore the overall contribution of the stress energy 
from causal seeds like defect fields,
because in general they are much smaller than
the total energy density of radiation and matter.
Thus in a FRW universe with only radiation
and matter components that evolve independently and adiabatically,
the scale factor $a(\eta)$ is determined 
by the unperturbed Einstein equation,
or equivalently the Friedmann equation:
\begin{equation}
  \label{fried}
  \dot{a}^2 + K a^2 =
  \frac{8\pi G \rho_{\rm m0} a_0^3}{3} (1+a) + \frac{\Lambda}{3}a^4,
\end{equation}
where
a dot represents a derivative with respect to the conformal time $\eta$,
$K$ is the curvature, $\rho_{\rm m}$ is the matter energy density,
$\Lambda$ is the cosmological constant,
and we have normalized $a_{\rm eq}=1$.
If we define
\begin{eqnarray}
  \Omega_{\rm m} &=& \frac{ 8\pi G \rho_{\rm m}}{3H^2}, \\ 
  \Omega_{\rm r} &=& \frac{ 8\pi G \rho_{\rm r}}{3H^2}=
  \frac{ 8\pi G \rho_{\rm m}}{3aH^2}, \\
  \Omega_\Lambda &=& \frac{ \Lambda}{3H^2},\\
  \Omega_K &=& -\frac{ K}{a^2H^2},
\end{eqnarray}
where $H=\dot{a}/a^2$ is the Hubble parameter,
then we have from (\ref{fried}) that
$\Omega_{\rm m}+\Omega_{\rm r}+\Omega_\Lambda+\Omega_K=1$ and
\begin{equation}
  \label{omega}
  \frac{\Omega_{\Lambda 0}}{\Omega_{\rm m0}} =
  \frac{\Lambda}{8\pi G \rho_{\rm m0}}, \quad
  \frac{\Omega_{K0}}{\Omega_{\rm m0}} =
  \frac{-3K}{8\pi G \rho_{\rm m0}a_0^2}.  
\end{equation}
We also notice that $\Omega_{\rm r0}/\Omega_{\rm m0}=a_0^{-1}\ll 1$.
We define
\begin{equation}
  \label{aABC}
  A=\frac{2(\sqrt{2}-1)}{\eta_{\rm eq}}\,, \quad
  B=\frac{\Omega_{K0}}{\Omega_{\rm m0}a_0}\,, \quad
  C=\frac{\Omega_{\Lambda 0}}{\Omega_{\rm m0}a_0^3}\,,
\end{equation}
where we note that $B,C\ll 1$ due to $a_0 \gg 1$ and 
$\Omega_{\rm m0} \sim \Omega_{K0} \sim \Omega_{\Lambda 0}$
according to the current observational results.
Thus we can rewrite equation~(\ref{fried}) as
\begin{equation}
  \label{eta_a}
  \left(\frac{da}{d\eta}\right)^2 =
  \bar{A}^2 ( 1 + a +  B a^2 + C a^4 )\,,
\end{equation}
where
\begin{equation}
  \label{barA}
  \bar{A} = \frac{1}{\eta_{\rm eq}}\int_0^1\frac{da}{(1+a+Ba^2+Ca^4)^{1/2}}
  \approx A\,.
\end{equation} 
Equation~(\ref{eta_a}) can then be numerically evaluated with certain choices of
$\Omega_{\rm m0}$, $\Omega_{\Lambda 0}$ and $\Omega_{K0}$.
Assuming three species of neutrinos and
using $\rho_{\gamma 0}=2.0747\times 10^{-51}$GeV$^4$ \cite{stdcos3} and
the fact that
at $\eta_{\rm eq}$ both the curvature and the cosmological constant
terms are negligible in (\ref{fried}),
we obtain
\begin{eqnarray}
  a_0 &=& 23219 \ \Omega_{\rm m0}h^2,
  \label{a_0}\\  
  \eta_{\rm eq} & =& 16.310\ (\Omega_{\rm m0}h^2)^{-1} {\rm Mpc}\,,
  \label{eta_eq}\\  
  t_{\rm eq} &=&3.4058 \times 10^{10} (\Omega_{\rm m0}h^2)^{-2}{\rm sec},
  \label{t_eq}
\end{eqnarray}
where $\eta_{\rm eq}$ is in the units measured today.
In certain cases, (\ref{eta_a}) can be exactly solved:
\begin{enumerate}
\item $K=\Lambda=0$ (i.e.\ $\Omega_{\rm m0}=1, \Omega_{\Lambda 0}=0$):
  \begin{eqnarray}
    a(\eta) & = & A^2\eta^2/4 + A\eta \,,
    \label{aflat} \\
    t(\eta) & = & A^2\eta^3/12 + A\eta^2/2 \,,
    \label{tflat}
  \end{eqnarray}
  which give $\eta_{\rm eq}=3t_{\rm eq}/\sqrt{2}$.
\item $K<0,\ \Lambda=0$ (i.e.\ $\Omega_{\rm m0}<1, \Omega_{\Lambda 0}=0$):
  \begin{eqnarray}
    a(\eta) & = &\frac{1}{2B}\left[
      \cosh{(\bar{A}\sqrt{B}\eta)}
      + 2\sqrt{B}\sinh{(\bar{A}\sqrt{B}\eta)} -1
    \right] \,,
    \label{aopen}    \\
    t(\eta) & = & \frac{1}{\bar{A}B}\left[
      \cosh{(\bar{A}\sqrt{B}\eta)}
      + \frac{1}{2\sqrt{B}}\sinh{(\bar{A}\sqrt{B}\eta)}
      - \frac{\bar{A}\eta}{2} - 1
    \right] \,.
    \label{topen}
  \end{eqnarray}
\item $K>0,\ \Lambda=0$ (i.e.\ $\Omega_{\rm m0}>1, \Omega_{\Lambda 0}=0$):
  \begin{eqnarray}
    a(\eta) & = & \frac{1}{2B}\left[
      \cos{(\bar{A}\sqrt{-B}\eta)}-2\sqrt{-B}\sin{(\bar{A}\sqrt{-B}\eta)}-1
    \right] \,,
    \label{aclose}\\
    t(\eta) & = & \frac{1}{\bar{A}B}\left[
      \cos{(\bar{A}\sqrt{-B}\eta)}
      + \frac{1}{2\sqrt{-B}}\sin{(\bar{A}\sqrt{-B}\eta)}
      - \frac{\bar{A}\eta}{2} -1
    \right] \,.
    \label{tclose}    
  \end{eqnarray}
\end{enumerate}
We notice that
at early times equations (\ref{aopen}--\ref{topen}) and 
(\ref{aclose}--\ref{tclose})
reduce to equations (\ref{aflat}--\ref{tflat}).
At late times
equations (\ref{aflat}--\ref{tflat}), (\ref{aopen}--\ref{topen}) 
and (\ref{aclose}--\ref{tclose}) give the asymptotic forms 
\begin{equation}
  a(\eta) \propto 
  \left\{
  \begin{array}{ll}
    \eta^2, & K=\Lambda=0,\\ 
    \exp(\bar{A}\sqrt{B}\eta), & K<0,\; \Lambda=0,\\
    1-\cos(\bar{A}\sqrt{-B}\eta), & K>0,\; \Lambda=0, 
  \end{array}
  \right.
\end{equation}
or
\begin{equation}
  a(t) \propto 
  \left\{
  \begin{array}{ll}
     t^{2/3}, & K=\Lambda=0,\\ 
     t,       & K<0,\; \Lambda=0,\\ 
     1-\cos[2\bar{A}(-B)^{3/2}t], & K>0,\; \Lambda=0.
  \end{array}
  \right.
\end{equation}
Figure~\ref{scalefactor} shows some examples of these solutions.
As we can see,
the destinies of universes in different cosmologies diverge,
although all have identical features around or before the radiation-matter
equality $t_{\rm eq}$.
This converging behavior at early times helps simplify
the calculation of cosmological perturbations with causal source,
since we know that this kind of perturbations 
are mainly contributed from the radiation-matter transition era.

\end{document}